\documentclass[%
reprint,preprintnumbers,
nofootinbib,
amsmath,amssymb,
aps
]{revtex4-1}
\pdfoutput=1
\usepackage{graphicx}
\usepackage[utf8]{inputenc}
\usepackage{flushend,comment}
\usepackage{dcolumn}
\usepackage{bm}
\usepackage{balance}
\usepackage[normalem]{ulem}
\usepackage[colorlinks = true,
            linkcolor = blue,
            urlcolor  = blue,
            citecolor = blue,
            anchorcolor = blue]{hyperref}
\usepackage{verbatim}
\usepackage{color,ulem}
\usepackage[english]{babel}
\usepackage{MnSymbol,wasysym,bbold}
\usepackage[utf8]{inputenc}
\input Starburst.fd
\newcommand*\initfamily{\usefont{U}{Starburst}{xl}{n}}\initfamily 

\newcommand{\beq}{\begin{eqnarray}}
\newcommand{\eeq}{\end{eqnarray}}
\usepackage{amsmath}
\usepackage{tikz}
\usetikzlibrary{decorations.pathmorphing}
\usetikzlibrary{shapes.misc}
\tikzset{cross/.style={cross out, draw=black, minimum size=8*(#1-\pgflinewidth), inner sep=0pt, outer sep=0pt},
cross/.default={1pt}}
\usetikzlibrary{patterns,math}

\def\be{\begin{equation}}
\def\ee{\end{equation}}
\def\bea{\begin{eqnarray}}
\def\eea{\end{eqnarray}}
\newcommand{\dd}{\mathrm{d}}
\definecolor{applegreen}{rgb}{0.55, 0.71, 0.0}

\begin{document}

\preprint{APCTP Pre2026 - 011, RIKEN-iTHEMS-Report-26, KUNS-3126}

\title{Deep learning emergent spacetime from fermionic spectral functions in holography}

\author{Koji Hashimoto$^{1}$}\email{koji@scphys.kyoto-u.ac.jp}
\author{Hyun-Sik Jeong$^{2,3}$}\email{hyunsik.jeong@apctp.org}
\author{Keun-Young Kim$^{4,5}$}\email{fortoe@gist.ac.kr}
\author{Daichi Takeda$^{6}$}\email{daichi.takeda@riken.jp}
\author{Kwan Yun$^{1,4}$\vspace{0.2cm}}\email{ludibriphy70@gm.gist.ac.kr}

\affiliation{$^{1}$Department of Physics, Kyoto University, Kyoto 606-8502, Japan}
\affiliation{$^{2}$Asia Pacific Center for Theoretical Physics, Pohang 37673, Korea}
\affiliation{$^{3}$Department of Physics, Pohang University of Science and Technology, Pohang 37673, Korea}
\affiliation{$^{4}$Department of Physics and Photon Science, Gwangju Institute of Science and Technology, 123 Cheomdan-gwagiro, Gwangju 61005, Korea}
\affiliation{$^{5}$Research Center for Photon Science Technology, Gwangju Institute of Science and Technology, 123 Cheomdan-gwagiro, Gwangju 61005, Korea}
\affiliation{$^{6}$iTHEMS, RIKEN, Wako, Saitama 351-0198, Japan}

\begin{abstract}
We present a physics-informed machine learning framework based on Neural Ordinary Differential Equations that solves the holographic inverse problem: reconstructing the bulk spacetime and gauge field of a charged AdS black hole directly from boundary fermionic spectral functions. Encoding the UV asymptotics, horizon regularity, and \textit{zero temperature} extremality as hard constraints in the neural network architecture, our framework reliably reconstructs the \textit{extremal} Reissner-Nordström AdS geometry across three quantum critical regimes set by the $U(1)$ probe charge---non-Fermi liquid, marginal Fermi liquid (strange metal), and Fermi-liquid-like states---and can jointly infer the probe charge itself to sub-percent accuracy. Relaxing the near-AdS boundary constraint uncovers a geometrical degeneracy: bulk profiles that differ throughout the radial direction but share the same near-horizon $AdS_2 \times \mathbb{R}^2$ data reproduce identical spectral functions near the Fermi surface. This isospectral non-uniqueness is precisely the bulk degeneracy expected on general holographic grounds at zero temperature, and its spontaneous emergence across independent training runs shows that the network isolates the IR CFT universality rather than overfitting a single UV completion.
\end{abstract}
\maketitle

%%%%%%%%%%%%%%%%%%%%%%%%%%%
%    
%%%%%%%%%%%%%%%%%%%%%%%%%%%
\section{Introduction}
One of the central problems in quantum condensed matter physics concerns the critical phenomena governing zero-temperature quantum phase transitions. At a quantum critical point (QCP), the divergence of the correlation length may give rise to scale invariance and emergent conformal symmetry~\cite{Sachdev:2011fcc}. However, exotic phases near a QCP---such as the strange metal regime in high-$T_c$ superconductors and heavy-fermion materials---depart fundamentally from Landau's Fermi liquid theory~\cite{Anderson:1990aa,Varma:1989aa,VARMA2002267}, exhibiting anomalous thermodynamic and transport properties. Understanding these non-Fermi liquid states requires analytical tools capable of handling strong correlations at finite density beyond weak-coupling field theories.
    
To address this challenge, holographic condensed matter theory (AdS/CMT) models certain classes of non-Fermi liquids and their associated quantum critical states by introducing probe Dirac fermions into the background of a charged AdS black hole~\cite{Lee:2008xf,Liu:2009dm,Cubrovic:2009ye,Faulkner:2010zz,Iqbal:2011ae}. This approach builds on the anti-de Sitter/conformal field theory (AdS/CFT) duality~\cite{Maldacena:1997re,Gubser:1998bc,Witten:1998qj}, which maps strongly coupled quantum many-body systems onto classical gravitational systems in higher dimension~\cite{Zaanen:2015oix,Ammon:2015wua,hartnoll2018}. In this setup, the charged black hole serves as the dual description of a strongly correlated ground state at finite density, and the scattering of Dirac fermion off the black hole horizon determines the boundary retarded Green's function $G_R(\omega, k)$ and the associated spectral function---a direct theoretical analog to angle-resolved photoemission spectroscopy (ARPES) measurements.

This holographic approach has offered novel insights into the low-energy dynamics of strongly interacting fermionic quantum critical state. Exploring different regions of the parameter space reveals both Fermi liquid-like~\cite{Cubrovic:2009ye} and non-Fermi-liquid behaviors~\cite{Liu:2009dm,Cubrovic:2009ye,Faulkner:2009wj}, showing that the infrared (IR) low-energy regime of these non-Fermi liquids is governed by a nontrivial quantum fixed point. Consequently, charged AdS black holes thus provide a robust and versatile tool for probing fermionic quantum criticality at finite density.

Historically, holographic modeling has operated in the forward direction: given a bulk gravity action, one solves the classical bulk equations of motion for a specified metric and matter fields, and integrates the probe field equations to compute boundary observables---a well-posed, essentially algorithmic procedure. The inverse problem is fundamentally different in character. Because the bulk fields are unknown continuous radial functions and the most boundary observables depend on them implicitly, through the non-linear solution of the bulk equation, there is often no closed-form procedure for inverting boundary data into a bulk profile. Recovering a holographic bulk configuration from a set of boundary measurements is therefore a non-trivial inverse problem, which may admit multiple, physically distinct bulk solutions consistent with the same boundary data.

Recently, physics-informed machine learning~\cite{Karniadakis_2021,Carleo:2019aa} has emerged as a powerful tool to bridge this gap in the context of holography (AdS/CFT)~\cite{Hashimoto:2018ftp,Tanaka_2021}. By identifying deep neural networks~\cite{LeCun_2015,SCHMIDHUBER201585} as continuous renormalization group flows, bulk geometries can be reconstructed from diverse boundary quantum data from strongly interacting QCD~\cite{Akutagawa:2020yeo,Akutagawa:2020yeo,Hashimoto:2020jug,Chen:2024ckb,Yan:2020wcd,Jeong:2025omu} and condensed matter systems~\cite{Li:2022zjc,Kim:2024car,Ahn:2024gjf,Ahn:2025tjp,Ahn:2024jkk,Xiao:2026dxe}. In particular, the Neural Ordinary Differential Equation (Neural ODE)~\cite{Chen:2018wjc,NEURIPS2019_21be9a4b,NEURIPS2020_293835c2,Yan2019OnRO} architecture allows neural networks to parameterize continuous bulk functions while preserving the exact differential equations of motion during optimization.

In this work, we present a Neural ODE-based inverse framework that reconstructs bulk spacetime metrics and gauge fields directly from the boundary fermionic spectral function, $\mathcal{A}(\omega, k):= {1}/{\pi}\, \text{Im} G_{R}(\omega,k)$. The physical motivation for this approach stems from the anomalous transport and spectral properties of strange metals, where low-energy excitations near the Fermi surface deviate sharply from standard Fermi-liquid quasiparticles~\cite{Varma:1989zz,Anderson:1990aa}. Such behavior points to an underlying quantum critical point at low energy scales~\cite{Aeppli_1997,Valla_1999,Tallon:1999aa,Marel_2003}---a phenomenon also prominently observed in heavy-fermion systems near a quantum phase transition~\cite{Gegenwart_2008}. While these features are historically modeled using phenomenological frameworks like the marginal Fermi liquid~\cite{Varma:1989zz}, holography offers an alternative geometric description.

To test whether bulk geometry can be reverse-engineered from these boundary quantum critical features, we apply our architecture to the zero-temperature ($T=0$) AdS background using non-Fermi liquid spectral data~\cite{Lee:2008xf,Liu:2009dm,Cubrovic:2009ye,Faulkner:2010zz}. We demonstrate that our deep learning model successfully recovers the emergent $AdS_2 \times \mathbb{R}^2$ quantum critical region from boundary fermionic input data. 

In Section~\ref{sec2}, we review the holographic Dirac equations for probe fermions in planar black hole spacetimes and establish the flow equations for computing retarded Green's functions. 
In Section~\ref{sec3}, we present the low-energy quantum critical behaviors of fermionic spectral functions generated from the extremal Reissner-Nordström AdS background, defining the three representative benchmark datasets. 
Section~\ref{sec4} details our physics-informed Neural ODE framework, including the hard-constrained neural network ansatz and the two-stage hybrid optimization strategy. 
In Section~\ref{sec5}, we demonstrate the successful reconstruction of the extremal Reissner-Nordström AdS spacetime and the joint determination of the probe fermion charge. 
In Section~\ref{sec5b}, we relax the near-boundary slope constraint, uncover an isospectral geometrical degeneracy among distinct bulk geometries sharing the same near-horizon $AdS_2\times\mathbb{R}^2$ spacetime.
We conclude with a summary in Section~\ref{sec6}.

%%%%%%%%%%%%%%%%%%%%%%%%%%%
%
%%%%%%%%%%%%%%%%%%%%%%%%%%%
\section{Holographic Dirac Equations}\label{sec2}
We consider a planar four-dimensional black brane background with the general metric ansatz:

\begin{equation}\label{eq:metric_ansatz}
\dd s^{2}=\frac{1}{z^{2}}\left[-f(z)\,\dd t^{2} + \frac{\dd z^{2}}{f(z)} + h(z) \, \dd \vec{x}_i^{2} \right] \,,
\end{equation}
where $z$ is the radial bulk coordinate, with the AdS boundary at $z\rightarrow0$ and the event horizon at $z_h=1$.

We introduce a probe Dirac fermion field $\Psi$ propagating in the background \eqref{eq:metric_ansatz}, with mass $m$ and $U(1)$ charge $q$---the latter identified with the charge of the dual boundary fermionic operator---governed by the curved-spacetime Dirac equation
\begin{equation}\label{eq:dirac_action}
\left(\Gamma^{M}D_{M} - m\right)\Psi = 0 \,,
\end{equation}
where the covariant derivative $D_M=\partial_M+\frac{1}{4}\omega_{AB,M}\Gamma^{AB}-iqA_M$ incorporates the spin connection $\omega_{AB,M}$ and the background $U(1)$ gauge potential $A_M = \left(A_t(z), 0, 0, 0\right)$. Here, $M$ denotes a bulk spacetime index, and $A,B$ denote tangent-space indices.
Gamma matrices can be expressed in the tangent space as $\Gamma^{M} = \Gamma^A e_A^M$ with the inverse vielbein $e_A^M$. We refer the reader to \cite{Liu:2009dm,Iqbal:2011ae} for a detailed review of the Dirac equations in holography.

By taking the Fourier transform along the boundary coordinates, we decompose the spinor as
\begin{equation}\label{eq:spinor_decomposition}
\Psi = \left(-g\,g^{zz}\right)^{-1/4}e^{i (kx-\omega t)}
\begin{pmatrix}
\phi_{+}(z)\\
\phi_{-}(z)
\end{pmatrix} \,,
\end{equation}
where, without loss of generality, spatial rotational symmetry in Eq.\eqref{eq:metric_ansatz} allows us to set $k_i=(k,0)$. Notably, this rescaling ansatz eliminates the spin connection terms.

To analyze the bulk Dirac equations, it is convenient to use the Gamma matrices $\Gamma^A$ as
\begin{equation}\label{eq:bulk_gamma_matrices}
\Gamma^{z}=
\begin{pmatrix}
-\mathbb{1}_{2} & 0\\
0 & \mathbb{1}_{2}
\end{pmatrix},
\qquad
\Gamma^{\mu}=
\begin{pmatrix}
0 & \gamma^{{\mu}}\\
\gamma^{{\mu}} & 0
\end{pmatrix} \,,
\end{equation}
where it can be further expressed with Pauli matrices $\sigma_i$ as $\gamma^{{t}}=i\sigma_{y},\, \gamma^{{x}}=\sigma_{x},$ and $\gamma^{{y}}=\sigma_{z}$.

Substituting the spinor decomposition \eqref{eq:spinor_decomposition} into the Dirac equation \eqref{eq:dirac_action} yields a coupled system of first-order radial equations for $y_{\pm}$ and $z_{\mp}$,
\begin{align}\label{eq:y_equation}
\begin{split}
&\sqrt{f(z)h(z)}\left(\partial_z\pm \frac{m}{z\sqrt{f(z)}} \right)y_{\pm}=\pm i\left(k-u\right)z_{\mp} \,,\\
&\sqrt{f(z)h(z)}\left(\partial_z\mp \frac{m}{z\sqrt{f(z)}} \right)z_{\mp}=\mp i\left(k+u\right)y_{\pm} \,,
\end{split}
\end{align}
where
\begin{equation}\label{eq:spinor_components}
\phi_{\pm} :=
\begin{pmatrix}
y_{\pm}\\
z_{\pm}
\end{pmatrix} \,,
\quad
u :=\sqrt{\frac{h(z)}{f(z)}}\left(\omega+qA_t(z)\right) \,.
\end{equation}

A convenient change of variables, defined by the ratios
\begin{equation}\label{eq:xi_definition}
\xi_{+}=i\frac{y_{-}}{z_{+}} \,,
\qquad
\xi_{-}=-i\frac{z_{-}}{y_{+}} \,,
\end{equation}
decouples the system \eqref{eq:y_equation} into two independent, nonlinear (Riccati-type) flow equations,
\begin{align}\label{eq:dirac_flow_fh_explicit}
\begin{split}
&\sqrt{f(z)}\,\partial_z\xi_{\pm} - \frac{2m}{z}\xi_{\pm} \mp \left[\frac{k}{\sqrt{h(z)}}\mp\frac{\omega+qA_t(z)}{\sqrt{f(z)}}\right] \\
&\pm\left[\frac{k}{\sqrt{h(z)}}\pm\frac{\omega+qA_t(z)}{\sqrt{f(z)}}\right]\xi_{\pm}^{2} = 0 \,.
\end{split}
\end{align}

These flow equations can be integrated from the horizon to the AdS boundary to compute the retarded Green's function $G_R(\omega,k)$ via holographic renormalization~\cite{Lee:2008xf,Liu:2009dm,Cubrovic:2009ye,Faulkner:2010zz,Iqbal:2011ae}
\begin{equation}\label{eq:boundary_green_function}
G_R=\lim_{z\rightarrow0}z^{-2m}
\begin{pmatrix}
\xi_{+}(z) & 0\\
0 & \xi_{-}(z)
\end{pmatrix} :=
\begin{pmatrix}
G_{+}(\omega,k) & 0\\
0 & G_{-}(\omega,k)
\end{pmatrix} \,.
\end{equation}

Notably, the two diagonal components are related by reversing the momentum sign~\cite{Liu:2009dm,Iqbal:2011ae}
\begin{equation}
G_{-}(\omega,k) = G_{+}(\omega,-k) \,,
\end{equation}
so that $G_{-}(\omega,k)$ alone already encodes the information carried by both components. The fermionic spectral function is given by the imaginary part of the retarded Green's function,
\begin{equation}\label{eq:spectral_function}
\mathcal{A}(\omega,k):=\frac{1}{\pi}\text{Im}G_{-}(\omega,k) \,.
\end{equation}

To solve the flow equation \eqref{eq:dirac_flow_fh_explicit}, an appropriate boundary condition must be imposed at the horizon. For $\omega\neq0$, demanding infalling boundary conditions yields
\begin{equation}\label{eq:infalling_condition}
\xi_{\pm}(z=1) = i \,.
\end{equation}
At zero temperature ($T=0$), however, the point $\omega=0$ requires special care: the extremal horizon produces a double zero in $f(z)$, causing the standard infalling condition to break down. For the extremal Reissner-Nordström AdS background, regularity at the horizon instead selects the zero-frequency condition as
\begin{equation}\label{eq:zero_frequency_condition}
\left.\xi_{\pm}\right|_{z=1,\,\omega=0} = \frac{m - \sqrt{k^{2} + m^{2} - \dfrac{(q\mu)^{2}}{6} - i\epsilon}}{\dfrac{q\mu}{\sqrt{6}} \pm k} \,, \quad \epsilon \rightarrow 0^{+} \,.
\end{equation}
Furthermore, near the double zero of $f(z)$ at the extremal horizon, $f\approx f_0(1-z)^2$, the Dirac field exhibits irregular singular behavior as $\phi_{\pm}(z) \approx e^{\frac{i \omega}{f_0(1-z)}}$. This singularity contrasts sharply with the finite-temperature case ($T\neq0$), where the near-horizon behavior is characterized by a power-law branch point, $(1-z)^{\frac{- i \omega}{4\pi T}}$.

%%%%%%%%%%%%%%%%%%%%%%%%%%%
%    
%%%%%%%%%%%%%%%%%%%%%%%%%%%
\section{Quantum critical behavior of spectral functions}\label{sec3}
To prepare the input fermionic spectral data for our neural network framework, we employ the low-energy boundary spectral functions generated from the extremal Reissner-Nordström AdS background, whose exact profiles correspond to 
\begin{align}\label{eq:rn_blackening}
\begin{split}
f(z) &=1-\left(1+\mu^{2}\right)z^{3}+\mu^{2}z^{4} \,, \quad
h(z) =1 \,, \\
A_{t}(z) &=\mu(1-z) \,,
\end{split}
\end{align}
where $\mu$ is the chemical potential, and Hawking temperature is given by $T={(3-\mu^2)}/{(4\pi)}$, thus the extremal limit ($T=0$) can be achieved by $\mu = \sqrt{3}$. In this limit the extremal geometry of \eqref{eq:rn_blackening} becomes $AdS_2 \times \mathbb{R}^2$ spacetime.

In this manuscript, we adopt the theoretical setup of Faulkner \textit{et al.}~\cite{Faulkner:2010zz}, in which a massless probe fermion with varying $U(1)$ charge $q$ models non-Fermi liquid and strange metal transport. A brief review of this setup is as follows.

Near the Fermi surface where $\omega \to 0$ and $k \to k_F$, the retarded Green's function $G_R(\omega, k)$ exhibits a characteristic low-energy quantum critical form governed by the emergent $AdS_2 \times \mathbb{R}^2$ near-horizon geometry:
\begin{equation}\label{eq:GR_fermi}
    G_R(\omega, k) = \frac{h_1}{k - k_F - \frac{\omega}{v_F} - \Sigma(\omega, k)} \,,
\end{equation}
where $k_F$ is the Fermi momentum, $v_F$ the Fermi velocity, $h_1$ a numerical constant, and $\Sigma(\omega, k)$ represents the self-energy for excitations near the Fermi surface, originating from the coupling to the IR CFT, i.e., $\Sigma(\omega, k) \approx \mathcal{G}_k(\omega)$, where $\mathcal{G}_k(\omega)$ is the IR retarded Green’s function, which can be computed analytically~\cite{Faulkner:2010zz,Faulkner:2009wj}.

The low-frequency scaling behavior of the self-energy near the Fermi momentum is controlled by the IR conformal dimension (or scaling exponent $\nu_k$), given by
\begin{equation}\label{eq:IR_exponent}
\Sigma(\omega, k) \approx c(k_F) \, \omega^{2\nu_{k_F}} \,, \quad \nu_k := \sqrt{m^2 L_2^2 + k^2\frac{L_2^2}{L_x^2}  - q^2 e_d^2 } \,,
\end{equation}
where $e_d^2 := L_2^4 \,A_t'(1)^2$. Here, $c(k)$ is a complex analytic function of $k$, while $L_2$ and $L_x$ denote the AdS$_2$ and $\mathbb{R}^2$ radii, respectively. $\nu_{k_F}$ serves as the dynamical critical exponent determining the lifetime and dispersion of low-energy excitations near $k_F$.

Depending on the magnitude of $\nu_{k_F}$, the spectral functions of massless fermions exhibit three representative quantum critical behaviors~\cite{Faulkner:2010zz}, which serve as distinct benchmark datasets for our machine learning training:
\begin{figure*}[]
    \centering
    \includegraphics[height=0.25\linewidth]{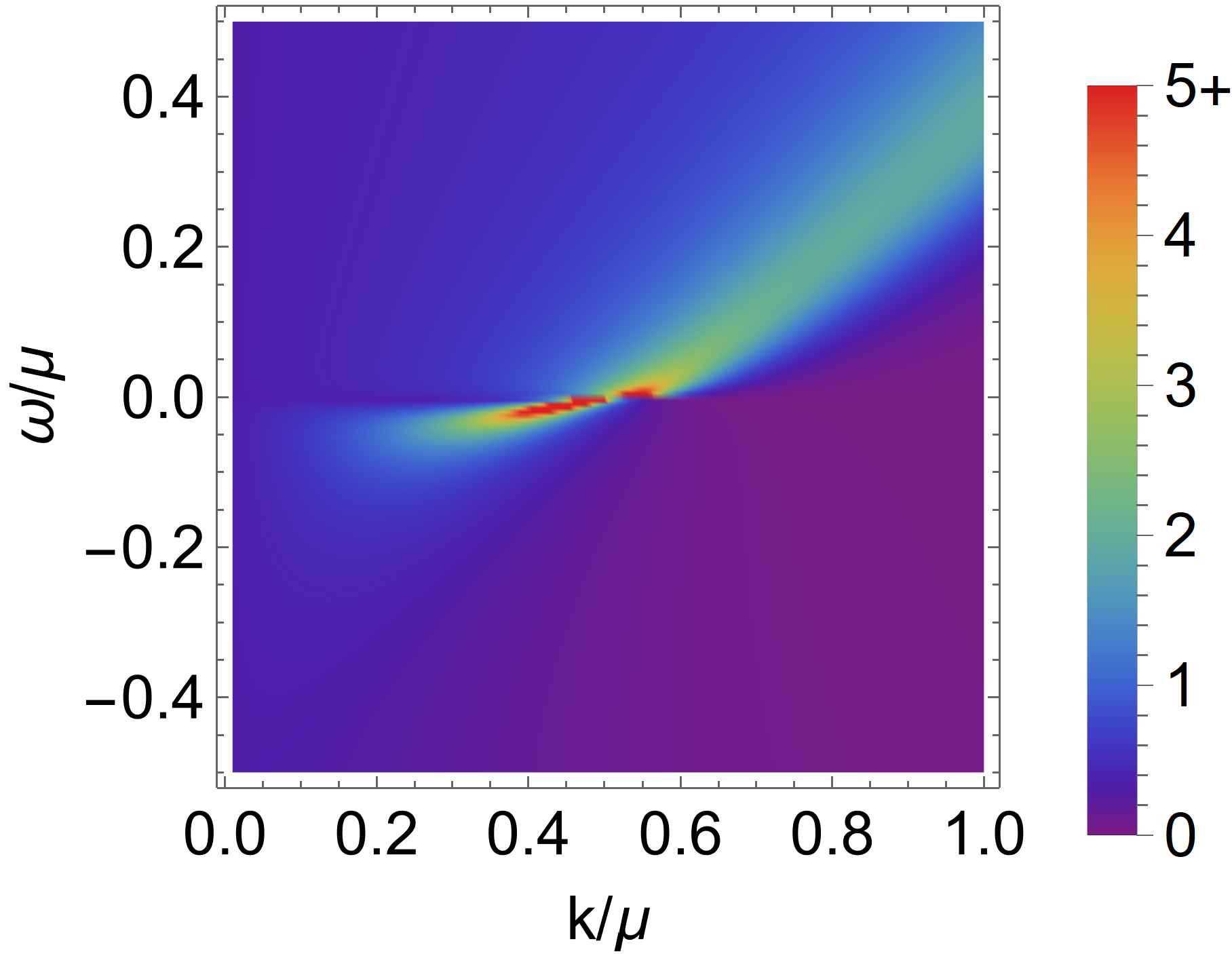}
    \includegraphics[height=0.25\linewidth]{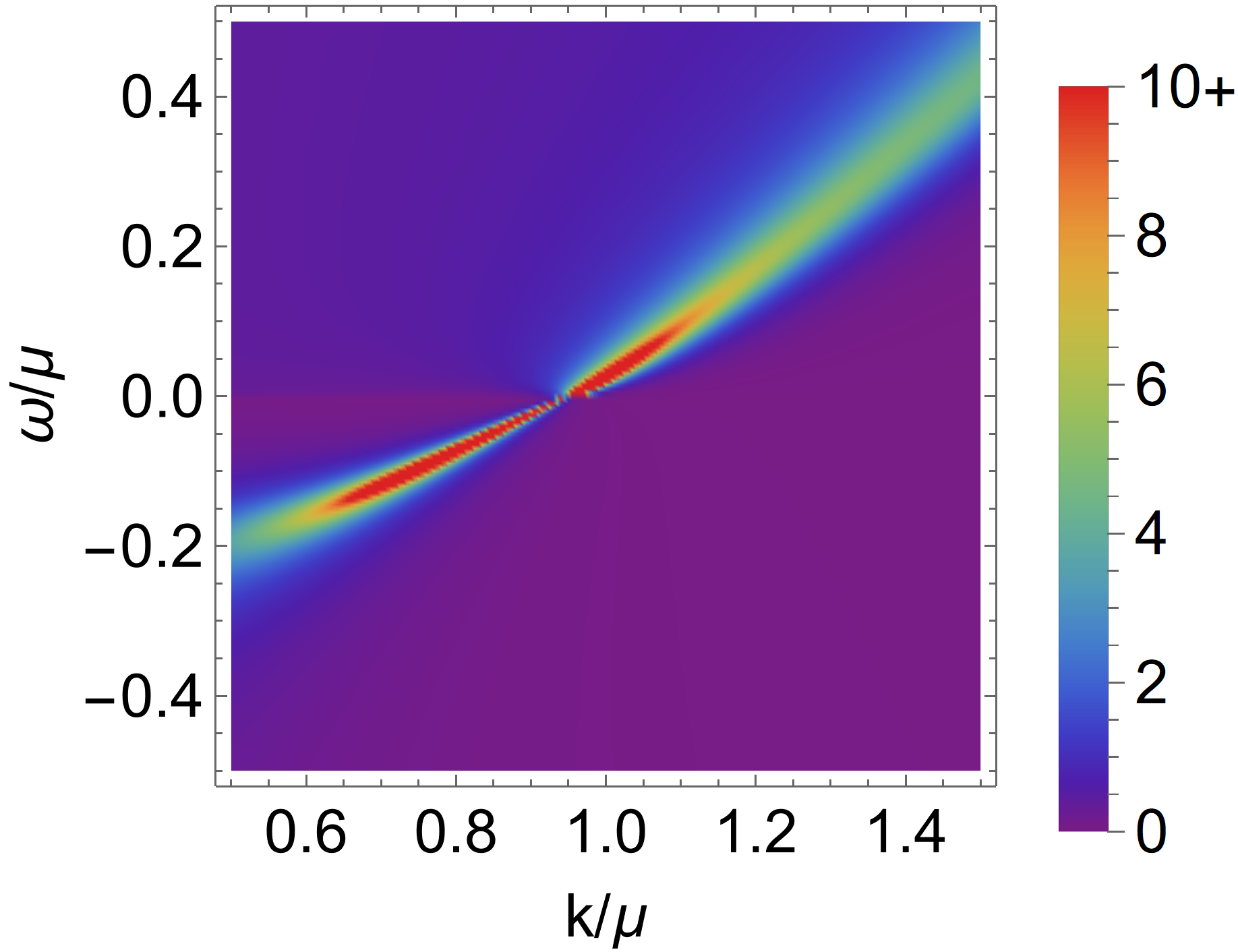}
    \includegraphics[height=0.25\linewidth]{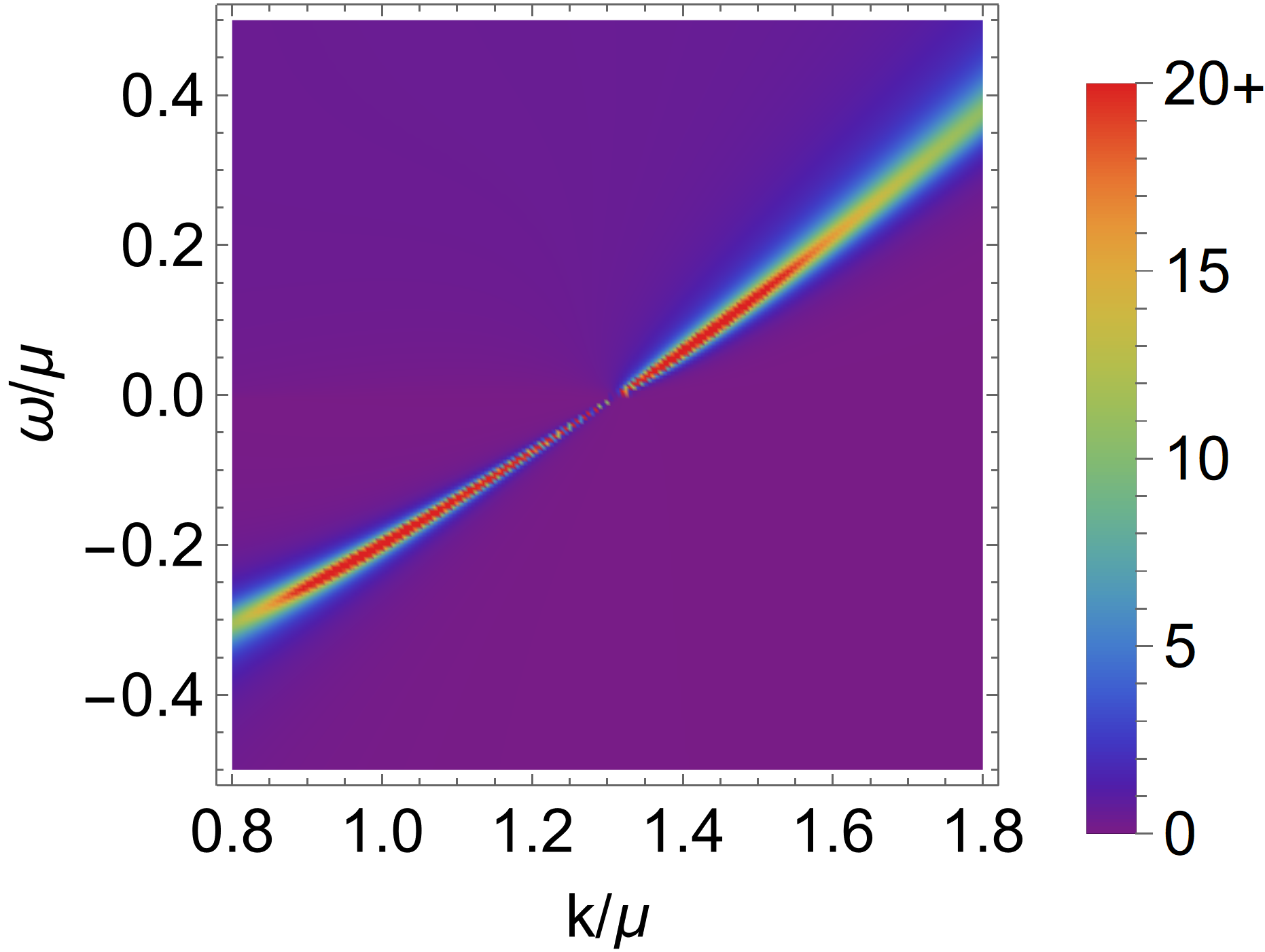}
\caption{The fermionic spectral function $\mathcal{A}(\omega,k) = \frac{1}{\pi}\text{Im}\,G_-(\omega,k)$ of the extremal Reissner-Nordström AdS black hole for three representative probe fermion charges: Data A ($q=1$, left), Data B ($q=1.56$, middle), and Data C ($q=2$, right). (a) Data A ($\nu_{k_F}=0.24 < 1/2$) displays a broad, incoherent peak with a non-linear dispersion $\omega_*(k) \approx (k-k_F)^{2.08}$ and a vanishing quasiparticle residue, representing a non-Fermi liquid without long-lived excitations. (b) Data B ($\nu_{k_F} = 1/2$) realizes the marginal Fermi liquid regime with a quasi-linear dispersion $\omega_*(k) \approx \frac{k-k_F}{\ln|k-k_F|}$ and a linear scattering rate $\Gamma \propto \omega$, forming a moderately broad spectral ridge characteristic of strange metal phenomenology. (c) Data C ($\nu_{k_F}=0.73 > 1/2$) exhibits an asymptotically linear dispersion $\omega_*(k) \approx v_F(k-k_F)$ with a vanishing relative width $\Gamma/\omega_* \to 0$ near $k_F$, producing a sharp, long-lived quasiparticle peak with a non-zero residue.}\label{fig:RN_spectral}
\end{figure*}

\begin{itemize}
\item \textbf{Data A ($\nu_{k_F} < 1/2$): Non-Fermi Liquid} \\
For $q = 1$, the Fermi momentum is $k_F = 0.53 \mu$, yielding $\nu_{k_F} = 0.24$ $(< 1/2)$. In this regime, the self-energy $\Sigma(\omega, k) \approx \omega^{2\nu_{k_F}}$ dominates over the linear analytic term $\omega/v_F$ in the denominator of Eq.~\eqref{eq:GR_fermi}. While a dispersion relation is still present, it becomes distinctly non-linear, $\omega_*(k) \approx (k - k_F)^{\frac{1}{2\nu_{k_F}}}$ (with $1/2\nu_{k_F} \approx 2.08$). Because the decay width ($\Gamma \propto \text{Im}\Sigma$) scales with the same power as the real part and remains comparable to the excitation energy, the spectral feature forms a broad, incoherent peak rather than a sharp excitation. The quasiparticle residue vanishes at the Fermi surface, representing a non-Fermi liquid devoid of long-lived quasiparticle excitations.

\item \textbf{Data B ($\nu_{k_F} = 1/2$): Marginal Fermi Liquid} \\
For $q = 1.56$, the Fermi surface lies at $k_F = 0.952 \mu$, corresponding precisely to the critical value $\nu_{k_F} = 1/2$. In this case, the linear frequency term $\omega/v_F$ and the IR self-energy share the same low-frequency scaling, giving rise to a characteristic logarithmic self-energy:
\begin{equation}
\text{Re} \Sigma(\omega) \approx \omega \log \omega \,, \quad 
\text{Im} \Sigma(\omega) \approx \omega \,.
\end{equation}
The real part produces a quasi-linear dispersion
$\omega_*(k) \approx \frac{k - k_F}{\ln |k - k_F|}$, ensuring a clear spectral ridge trajectory. However, this leads to a single-particle scattering rate $\Gamma$ that depends linearly on frequency, causing the quasiparticle residue to vanish logarithmically as $k \to k_F$. This directly realizes the marginal Fermi liquid state~\cite{Varma:1989zz} characteristic of strange metal phenomenology and ARPES observations on cuprates~\cite{Abrahams_2000}. Consequently, the spectral peak exhibits an intermediate broadness: it is neither a sharp, long-lived quasiparticle nor a featureless continuum, but a well-defined yet broad spectral ridge.

\item \textbf{Data C ($\nu_{k_F} > 1/2$): Fermi-Liquid-Like} \\
For $q = 2$, one obtains $k_F = 1.315 \mu$ with a conformal dimension $\nu_{k_F} = 0.73$ $(> 1/2)$. In this case, the analytic linear term $\omega/v_F$ dominates the real part of the denominator of \eqref{eq:GR_fermi}, restoring an asymptotically linear dispersion $\omega_*(k) \approx v_F (k - k_F)$ as in a Fermi liquid. Nevertheless, the scattering rate scales differently from that of a Fermi liquid. The relative spectral width scales as $\Gamma(k) / |\omega_*(k)| \approx |k - k_F|^{2\nu_{k_F}-1} \to 0$ as $k \to k_F$. This suppression of the scattering rate sharpens the peak dramatically, giving rise to long-lived, sharp quasiparticle-like excitations with a non-zero residue at the Fermi surface.
\end{itemize}

We display the fermionic spectral functions~\eqref{eq:spectral_function} of the extremal Reissner-Nordström black hole for $q = 1, 1.56,$ and $2$ in Fig.~\ref{fig:RN_spectral}, which reproduce the results in Ref.~\cite{Faulkner:2010zz} and clearly demonstrate the physical descriptions above. 

In our machine learning framework, these three distinct spectral regimes---ranging from an incoherent non-Fermi liquid peak to a sharp Fermi-liquid-like quasiparticle excitation---serve as three separate boundary input datasets on which the inverse model is trained independently.

%%%%%%%%%%%%%%%%%%%%%%%%%%%
%    
%%%%%%%%%%%%%%%%%%%%%%%%%%%
\section{Neural ODE Framework}\label{sec4}
To address the holographic inverse problem, we construct a Neural Ordinary Differential Equation (Neural ODE) framework~\cite{Chen:2018wjc,NEURIPS2019_21be9a4b,NEURIPS2020_293835c2,Yan2019OnRO} that directly reconstructs the unknown continuous bulk profile $\{f(z), h(z), A_t(z)\}$ from boundary fermionic spectral data. While the forward holographic calculation determines boundary observables from a specified bulk background, our framework optimizes neural network representations of the bulk metric and gauge fields by embedding the exact radial Dirac flow equation Eq.\eqref{eq:dirac_flow_fh_explicit} directly into the network's learning process.
\vspace{0.1cm}

\noindent{\textbf{Hard-Constrained Neural Network Ansatz.}}
Instead of learning unconstrained bulk functions, we incorporate the required physical boundary conditions (AdS boundary at $z \to 0$), zero temperature ($f'(1)= 0$), and black brane conditions ($f(1)=0$) directly into the network architecture through hard constraints:
\begin{align}\label{eq:nn_ansatz_At}
\begin{split}
    f_\theta(z) &= (1-z)^2 \left[ 1 + (a+2)z + z^2 D_f(z;\theta_f) \right] \,, \\
    h_\theta(z) &= 1 + az + z^2 D_h(z;\theta_h) \,, \\
    A_{t,\theta}(z) &= \mu(1-z) + z(1-z) D_A(z;\theta_A) \,,
\end{split}
\end{align}
where $D_f(z;\theta_f)$, $D_h(z;\theta_h)$, and $D_A(z;\theta_A)$ are three independent scalar Multi-Layer Perceptrons (MLPs) parameterized by trainable weights $\theta = \{\theta_f, \theta_h, \theta_A\}$. Each MLP consists of 3 hidden layers with 20 neurons per layer and utilizes the Softplus activation function.

By construction, Eq.~\eqref{eq:nn_ansatz_At} guarantees the UV AdS boundary conditions $f_\theta(0) = h_\theta(0) = 1$, as well as the IR horizon regularity $f_\theta(1) = 0$ and $A_{t,\theta}(1) = 0$. The double zero $(1-z)^2$ in $f_\theta(z)$ strictly enforces zero Hawking temperature ($T = 0$). Furthermore, for computational convenience, we also enforce the near-boundary derivative conditions $f_\theta'(0) = h_\theta'(0) = a$, where $a$ is a common asymptotic slope parameter, which vanishes for Reissner-Nordström AdS background \eqref{eq:rn_blackening}.
\vspace{0.1cm}
\begin{figure*}[]
    \centering
\includegraphics[width=0.3\linewidth]{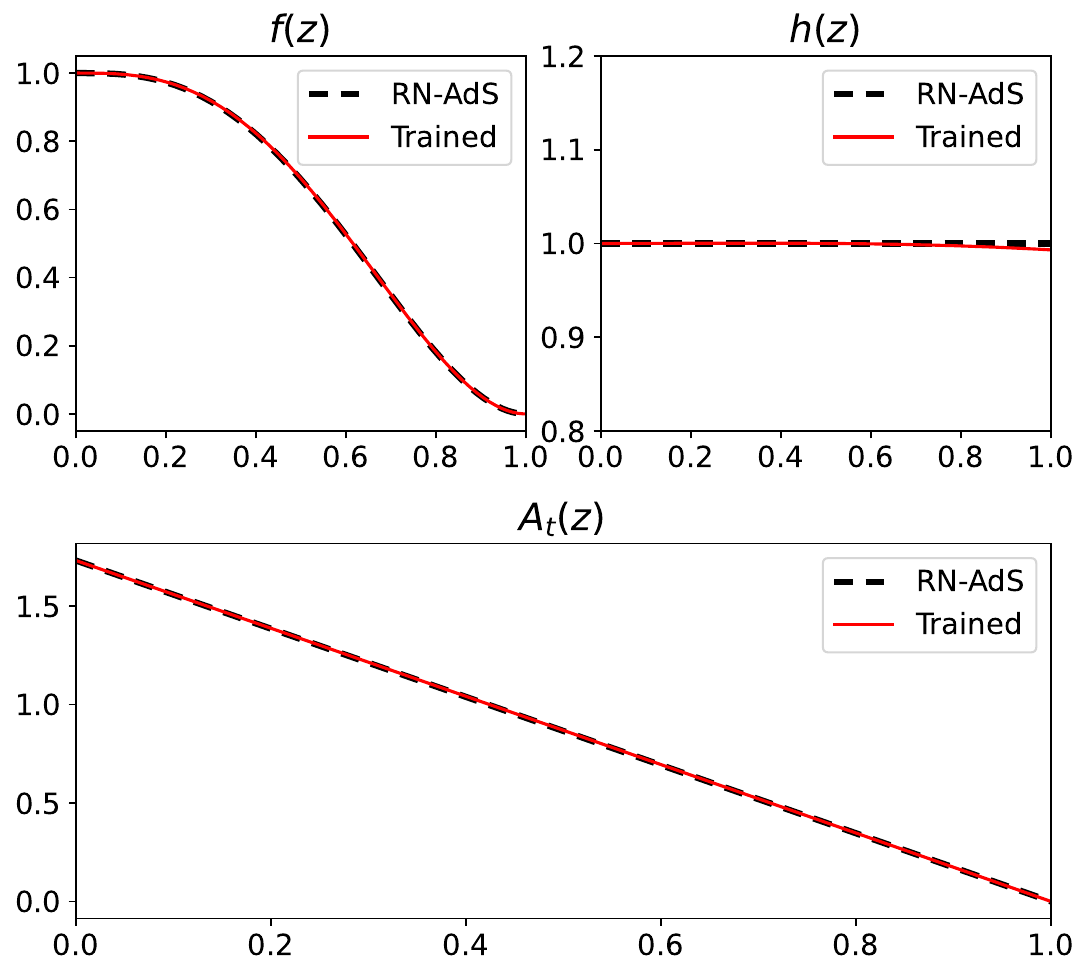}
\qquad
\includegraphics[width=0.3\linewidth]{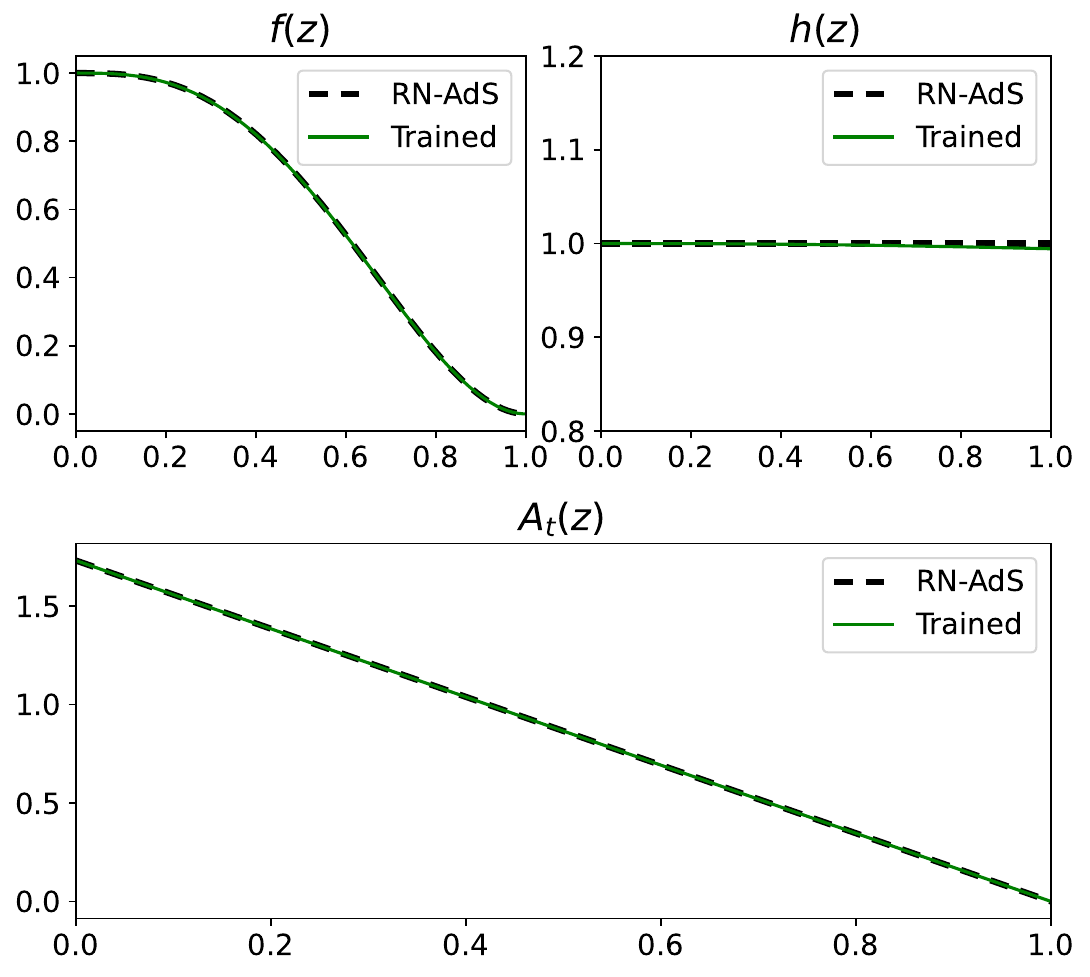}
\qquad
\includegraphics[width=0.3\linewidth]{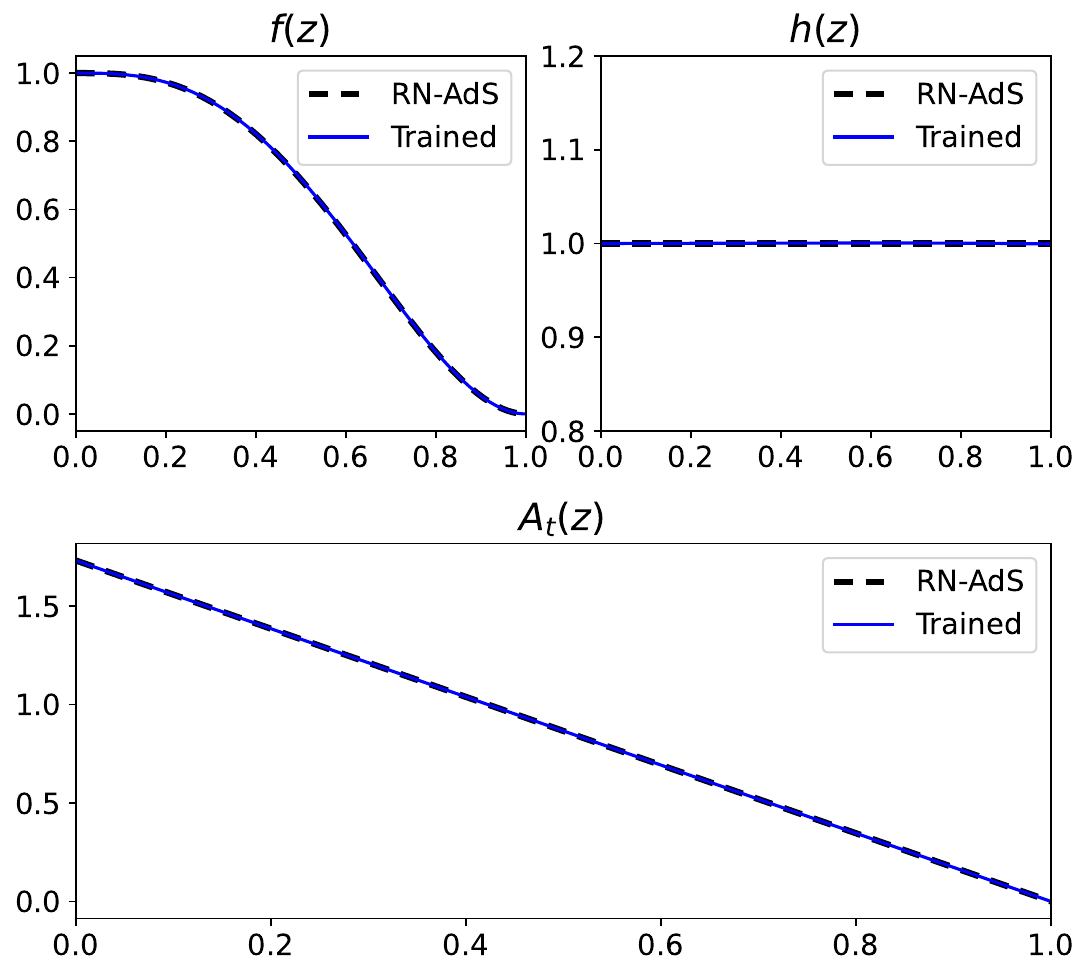}
    
\includegraphics[width=0.3\linewidth]{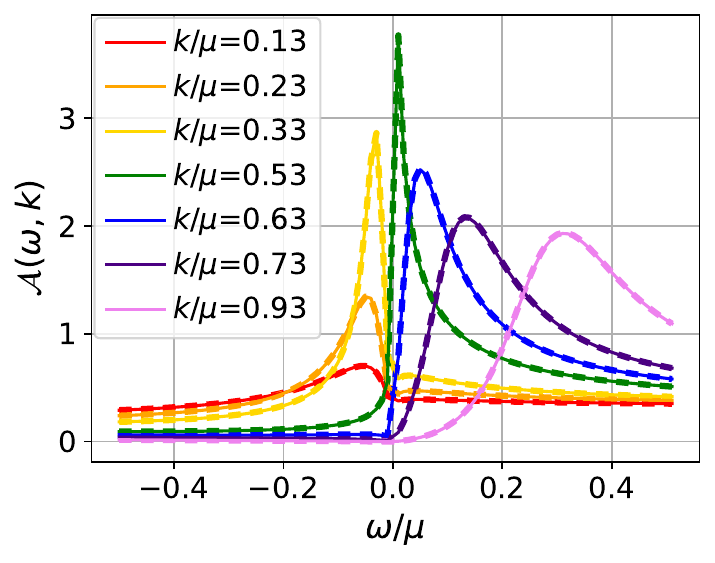}
\qquad
\includegraphics[width=0.3\linewidth]{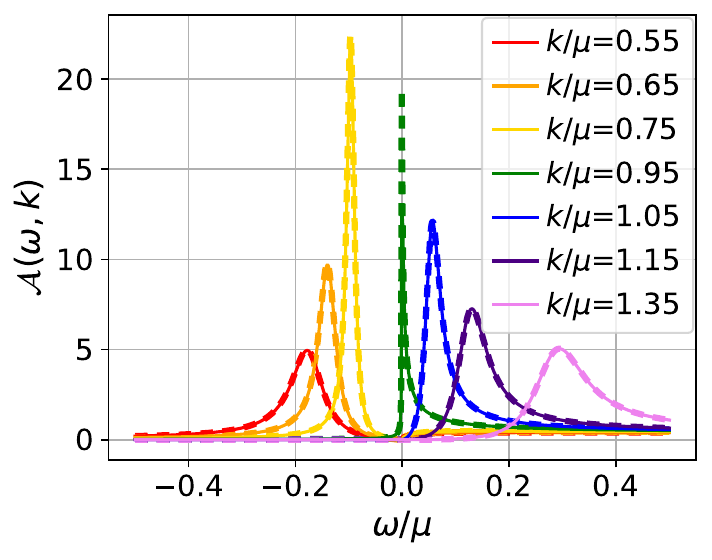}
\qquad
\includegraphics[width=0.3\linewidth]{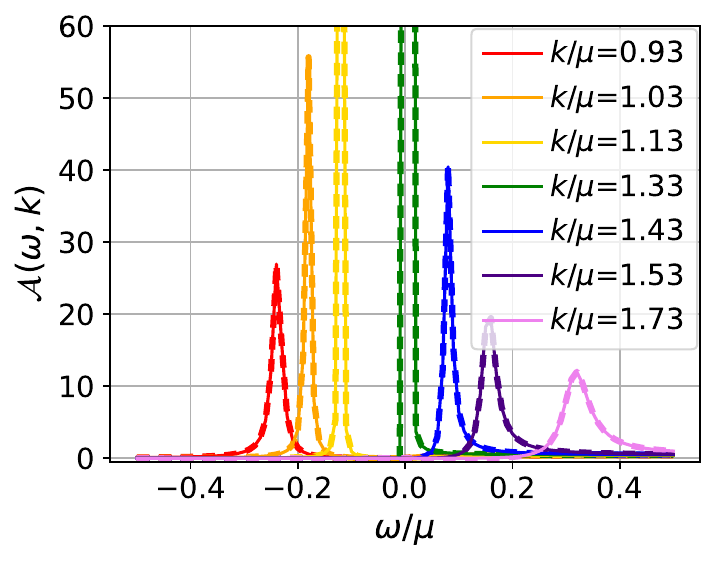}
\caption{Reconstructed bulk profiles $\{f(z), h(z), A_t(z)\}$ obtained from the quantum critical fermionic spectral data shown in Fig.~\ref{fig:RN_spectral}: Data A ($q=1$, left), Data B ($q=1.56$, middle), and Data C ($q=2$, right). Upper panels: The trained neural network profiles (colored curves) closely match the exact extremal Reissner-Nordström AdS black hole solutions \eqref{eq:rn_blackening} with $\mu = \sqrt{3}$ (dashed lines). Lower panels: The corresponding reconstructed fermionic spectral functions $\mathcal{A}_\theta(\omega_i, k_i)$ (solid lines) compared against the input spectral function $\mathcal{A}^{\mathrm{data}}_i$ (dashed lines). Across all three datasets, the reconstructed bulk profiles accurately reproduce even the out-of-sample test data, represented by the orange and blue curves.}\label{fig:RN_trained}
\end{figure*}

\noindent{\textbf{Numerical Integration and Optimization Strategy.}}
For each boundary data point $(\omega_i, k_i)$ sampled on a regular grid from the spectral functions Eq.\eqref{eq:spectral_function} (or Fig.~\ref{fig:RN_spectral}), the dynamical spinor ratio $\xi_-(z)$ is integrated along the radial direction using the Dirac flow equation \eqref{eq:dirac_flow_fh_explicit}. The training data are discretized on a uniform $12\times5$ grid in ($\omega/\mu\times k/\mu$), giving $N_{\rm data}=60$ points per dataset. We choose a small frequency window,
$\omega/\mu\in[-0.45,0.45]$, for all three datasets so as to cover the IR Green's function, while the momentum range is centered on the respective Fermi momentum: $k/\mu\in[0.13,0.93]$, $[0.55,1.35]$, and $[0.93,1.73]$ for Data A, B, and C, respectively.

The integration proceeds from the near-horizon IR cutoff $z_h = 1 - \epsilon_h$ to the near-boundary UV cutoff $z_b = \epsilon_b$, with $\epsilon_h = 5 \times 10^{-3}$ and $\epsilon_b = 10^{-4}$. We employ the Tsitouras fifth-order Runge-Kutta method (Tsit5) with an ODE error tolerance of $10^{-6}$. The deep learning-induced spectral function is obtained directly from the boundary value: 
\begin{equation}
\mathcal{A}_\theta(\omega_i, k_i) = \frac{1}{\pi} \text{Im}\,\xi_-(z_b) \,.
\end{equation}

The network parameters $\theta$ are trained by minimizing the unweighted mean-squared error (MSE) loss between the $\mathcal{A}_\theta(\omega_i, k_i)$ and the boundary training data $\mathcal{A}^{\mathrm{data}}_i$:
\begin{equation}\label{eq:spectral_loss}
    \mathcal{L}_{\mathrm{data}}(\theta) = \frac{1}{N_{\mathrm{data}}} \sum_{i=1}^{N_{\mathrm{data}}} \left[ \mathcal{A}_\theta(\omega_i, k_i) - \mathcal{A}^{\mathrm{data}}_i \right]^2 \,.
\end{equation}
Because all physical endpoint and horizon conditions are built identically into the ansatz \eqref{eq:nn_ansatz_At}, no additional penalty terms or regularization loss components are required.

To achieve both fast global exploration and high-precision convergence, we implement a two-stage hybrid optimization strategy:
(I) Adam~\cite{Kingma:2014vow}: We first train using the Adam optimizer with a learning rate $\eta_{\mathrm{Adam}} = 10^{-3}$ until the loss drops to $\mathcal{O}(10^{-1})$ or $\mathcal{O}(10^0)$.
(II) L-BFGS~\cite{Liu:1989esw}: Optimization is then transitioned to the L-BFGS algorithm ($\eta_{\mathrm{L-BFGS}} = 0.1$, \texttt{max\_iter} = 5, \texttt{tolerance\_grad} = $10^{-5}$, \texttt{tolerance\_change} = $10^{-5}$) with the strong-Wolfe line search.

Training is terminated when the standard deviation of the loss over the most recent 10 epochs falls below $\operatorname{Std}\left( \mathcal{L}_{n-9}, \ldots, \mathcal{L}_n \right) < 10^{-9}$.

\begin{figure*}[t]
    \centering
    \includegraphics[width=0.28\linewidth]{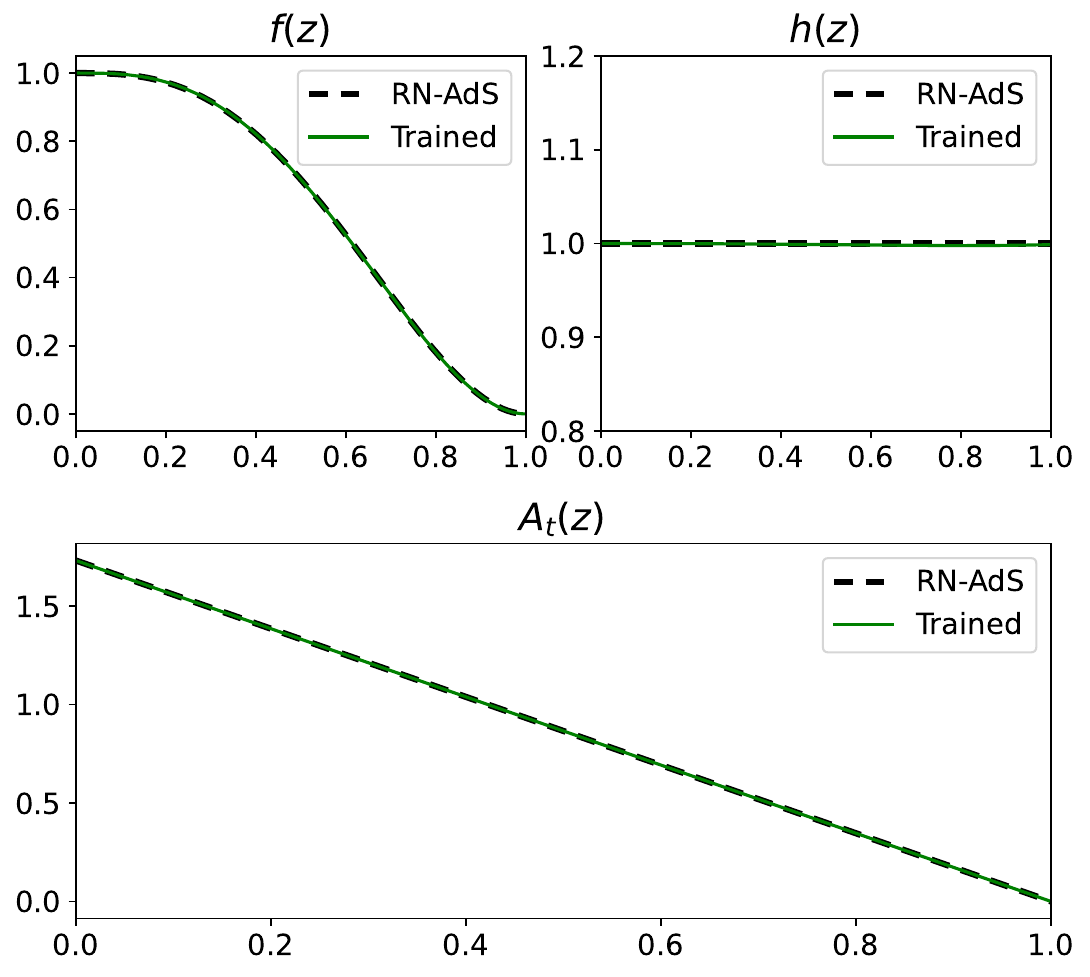}
\qquad
    \includegraphics[width=0.31\linewidth]{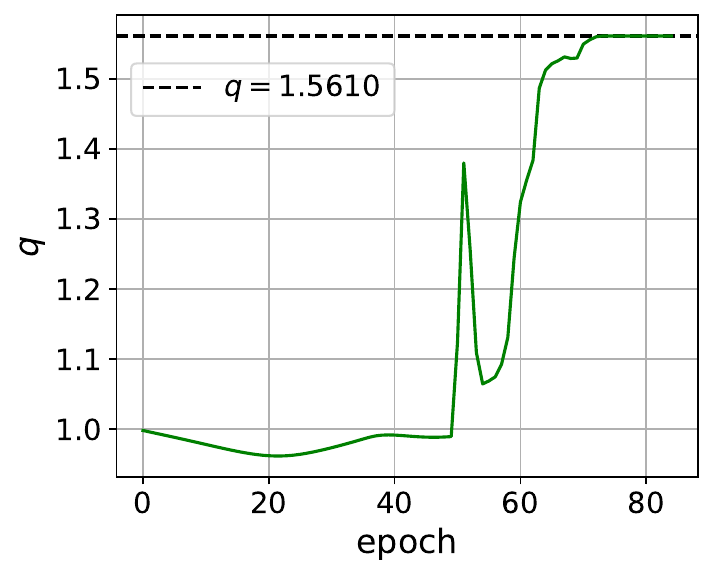}
\qquad
    \includegraphics[width=0.32\linewidth]{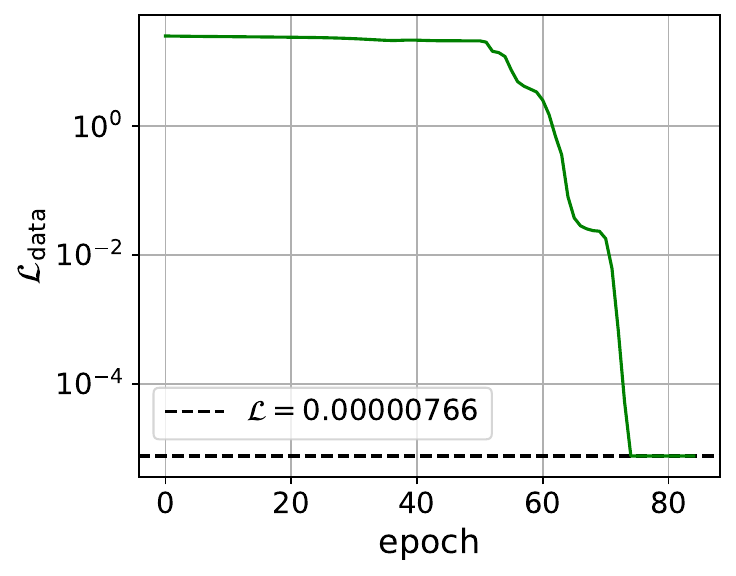}
    \caption{Simultaneous reconstruction of bulk profiles and probe fermion charge $q$ trained on Data B (marginal Fermi liquid). Left: Reconstructed bulk metric and gauge fields compared against the target extremal Reissner-Nordström AdS background. Middle: Optimization trajectory of the trainable charge $q$, converging to $q_{\mathrm{trained}} = 1.5610$ (target $q = 1.56$). Right: Evolution of the training loss during the hybrid Adam and L-BFGS optimization, reaching a final loss of $\mathcal{L} = 7.66 \times 10^{-6}$.}\label{fig:RN_trained_q}
\end{figure*}

\begin{figure*}[]
    \centering
    \includegraphics[width=0.28\linewidth]{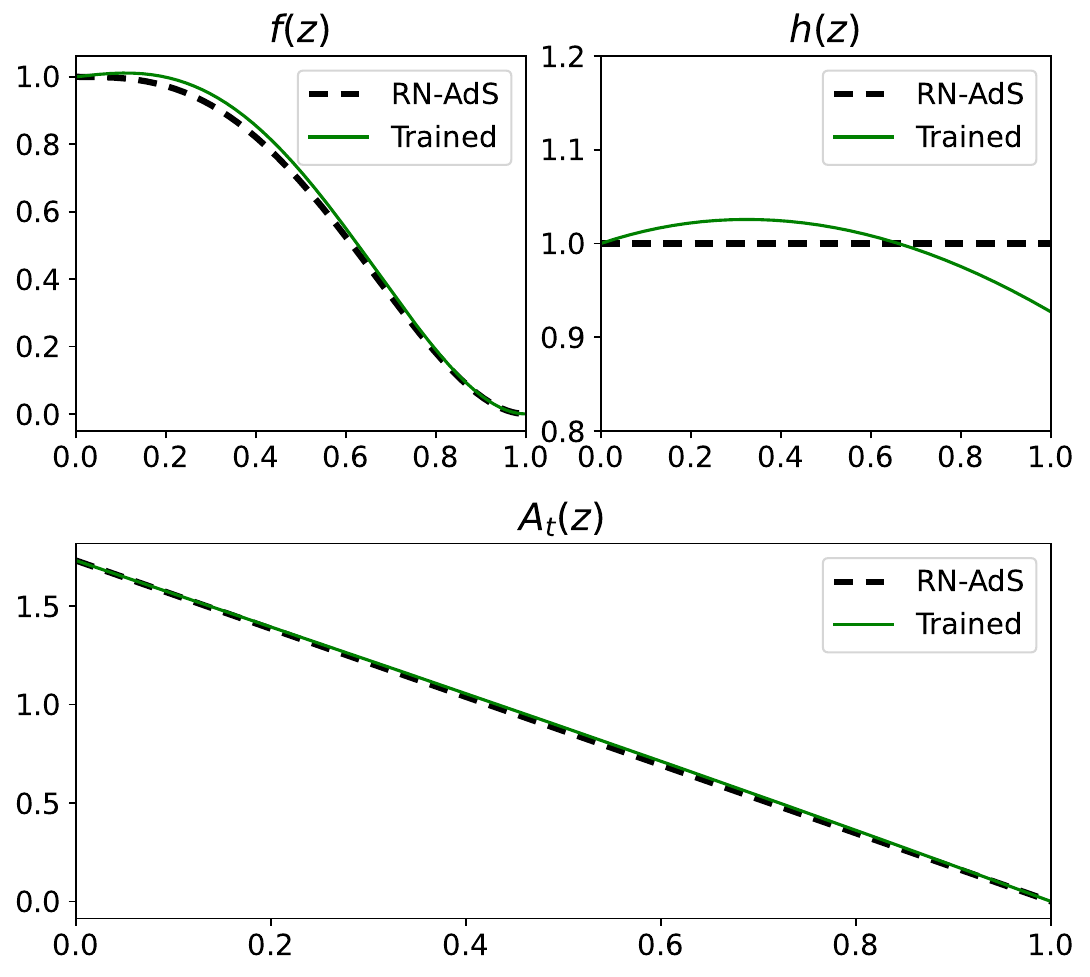}
\qquad
    \includegraphics[width=0.31\linewidth]{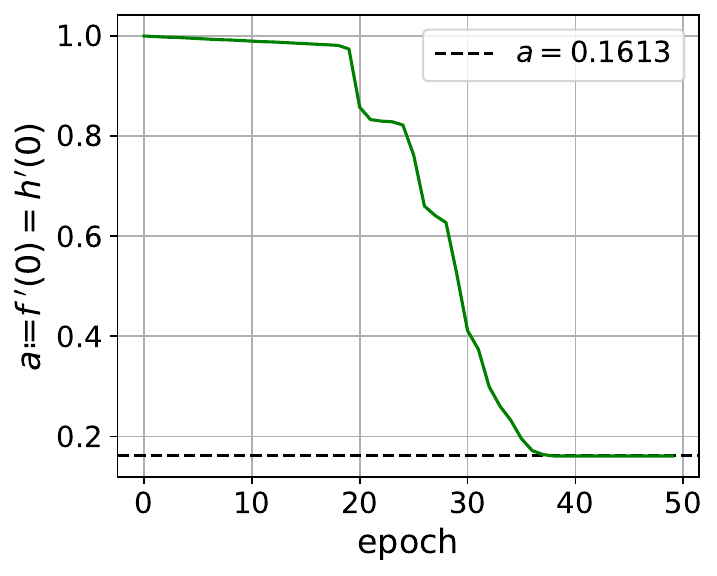}
\qquad
    \includegraphics[width=0.32\linewidth]{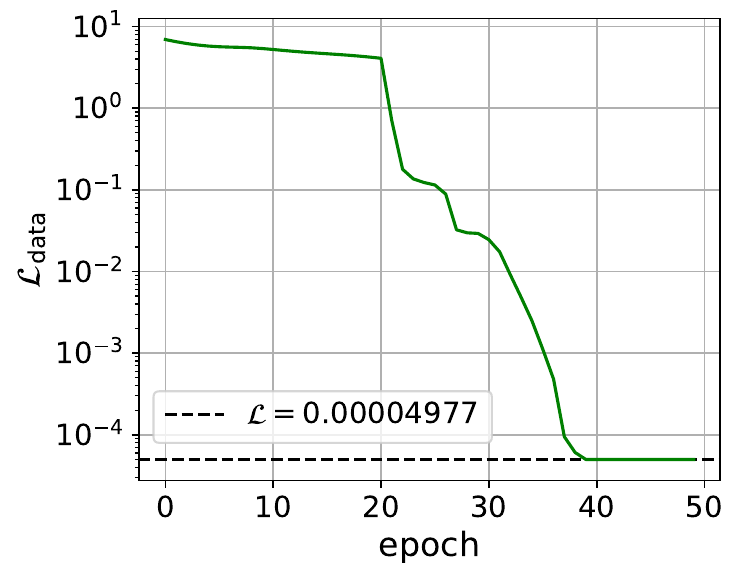}
\caption{The trained results using Data B ($q=1.56$) when relaxing the near-boundary derivative condition $a = f'(0) = h'(0)$. Left: The trained bulk metric functions $f(z)$, $h(z)$ and gauge potential $A_t(z)$ (green solid lines) compared with the exact extremal Reissner-Nordström AdS background (black dashed lines). Middle: Evolution of the trainable slope parameter $a$ during training, converging to $a_{\mathrm{trained}} \approx 0.1613$. Right: Convergence of the training loss $\mathcal{L}_{\mathrm{data}}$ as a function of epoch, reaching a final loss $\approx 4.98 \times 10^{-5}$.}\label{fig:RN_trained_a}
\end{figure*}

\begin{figure*}[t]
    \centering
    \includegraphics[width=1\textwidth]{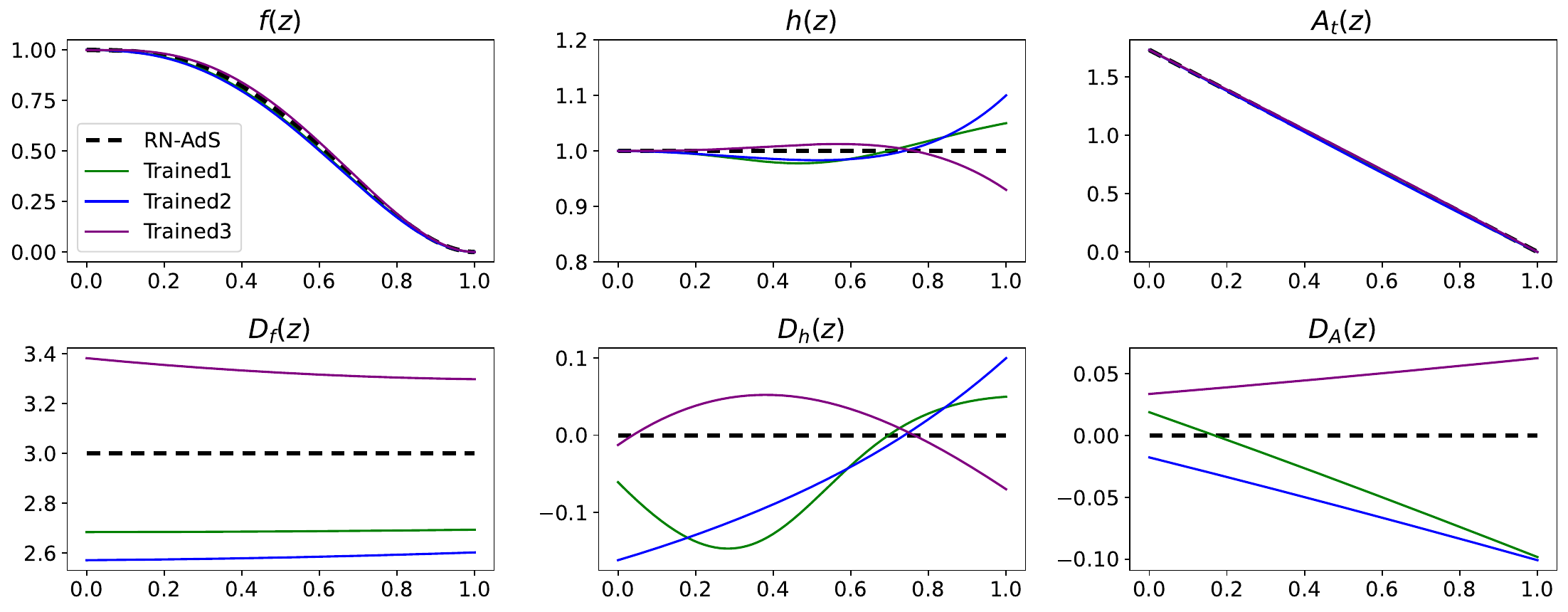}
\caption{The trained bulk profiles (top) and the functions $D_{f,h,A}(z)$ (bottom) obtained with the Neural ODE at $a=0$. The solid lines (green, blue, and purple) represent three different trained solutions, while the black dashed lines denote the exact extremal Reissner-Nordström AdS solution. The three solutions differ substantially from the Reissner-Nordström AdS solution, closely reproducing the same spectral function data, with residual losses as $\mathcal{L}_{\rm data}=1.65\times10^{-5}$, $5.92\times10^{-4}$, and $5.05\times10^{-4}$, respectively.}\label{fig:Fixed_a}
\end{figure*}

%%%%%%%%%%%%%%%%%%%%%%%%%%%
%
%%%%%%%%%%%%%%%%%%%%%%%%%%%
\section{Deep learning emergent spacetime from fermionic spectral functions}\label{sec5}
\noindent{\textbf{Reconstruction of the extremal RN AdS spacetime.}}
We now present the bulk spacetime reconstruction using the quantum critical fermionic spectral data introduced in Fig.~\ref{fig:RN_spectral}. We begin with the baseline configuration: fixing the common asymptotic derivative parameter to $a=0$ in the neural network ansatz \eqref{eq:nn_ansatz_At} while setting the probe fermion charge $q$ to its exact values ($q=1, 1.56$, and $2$ for Data A, B, and C, respectively).

The trained bulk metric components $\{f(z), h(z)\}$ and gauge potential profile $A_t(z)$ are displayed in Fig.~\ref{fig:RN_trained}. Across all three probe charge regimes, the trained neural profiles match the target analytical extremal Reissner-Nordström (RN) AdS black hole geometry \eqref{eq:rn_blackening} with chemical potential $\mu=\sqrt{3}$. The optimization achieves final L-BFGS loss values of $4.65\times10^{-5}$, $1.78\times10^{-5}$, and $8.72\times10^{-3}$ for Data A, Data B, and Data C, respectively; the comparatively larger residual for Data C reflects the added numerical difficulty of resolving its sharp, long-lived quasiparticle peak on the same $12\times5$ discretization grid used for all three datasets. This high reconstruction fidelity is further demonstrated by the close agreement between the trained spectral functions $\mathcal{A}_\theta(\omega_i, k_i)$ and the target input data $\mathcal{A}^{\mathrm{data}}_i$, as shown in the lower panels of Fig.~\ref{fig:RN_trained}.
\vspace{0.1cm}

\noindent{\textbf{Reconstruction of bulk fields and fermion charge.}}
Next, we investigate whether a $U(1)$ fermion charge $q$ can also be identified simultaneously alongside the bulk fields from the given spectral functions. We demonstrate this joint optimization using Data B, corresponding to the marginal Fermi liquid (strange metal) case. Retaining the fixed derivative condition $a=0$, we promote the probe fermion charge $q$ to an independent trainable parameter optimized in addition to the network weights $\theta$ of bulk fields.

As shown in Fig.~\ref{fig:RN_trained_q}, the learned bulk fields again converge to the exact extremal RN AdS profile. Concurrently, the probe charge converges to
\begin{equation}
    q_{\mathrm{trained}} = 1.5610,
\end{equation}
which exhibits agreement with the reference true value $q = 1.56$ used to generate Data B, yielding a relative error of only $0.06\%$. The optimization achieves a final L-BFGS loss of $7.66 \times 10^{-6}$. This result demonstrates that boundary fermionic spectral data encode sufficient physical information to reconstruct not only the continuous bulk spacetime and gauge potential but also the intrinsic charge of the probe fermionic operator.

%%%%%%%%%%%%%%%%%%%%%%%%%%%
%
%%%%%%%%%%%%%%%%%%%%%%%%%%%
\section{Isospectral Bulk Spacetime and IR Universality}\label{sec5b}
\noindent{\textbf{Relaxing the near-boundary slope and degenerate spacetime.}}
We investigate the role of the near-boundary derivative constraint $f'(0) = h'(0) = a$ imposed in the preceding reconstructions. To examine whether the boundary spectral data uniquely specifies the full bulk metric profile, we train the model on Data B ($q=1.56$) with the bulk fermion charge $q$ fixed, but promote the common asymptotic slope $a$ in Eq.~\eqref{eq:nn_ansatz_At} to an independent trainable scalar parameter.

Upon optimization, the slope parameter converges to 
\begin{equation}
    a_{\mathrm{trained}} \approx 0.1613 \,,
\end{equation}
with the corresponding bulk profiles displayed in Fig.~\ref{fig:RN_trained_a}. Despite departing visibly from the exact extremal RN AdS solution (which corresponds to $a=0$), these alternative profiles reproduce the input Data B spectrum with a remarkably low final loss of $\mathcal{L}_{\mathrm{data}} = 4.98 \times 10^{-5}$. This finding demonstrates an intriguing geometrical ``degeneracy": distinct bulk geometries across the intermediate radial regime can generate the same boundary fermionic spectral functions. This non-uniqueness is reflected directly in the optimization itself: independent training runs converge to different values of $a$, each giving rise to a different, yet equally consistent, degenerate bulk profile.
\vspace{0.1cm}

\noindent{\textbf{Revisiting the case of $a=0$.}}
Notably, the constraint $a=0$ alone does not always guarantee RN AdS spacetime: whereas the baseline reconstructions of Sec.~\ref{sec5} (Fig.~\ref{fig:RN_trained}) reliably converge to the exact RN AdS profile from generic random initializations of the network weights, we find that specific, atypical initial conditions---such as oscillatory initial bulk profiles---can occasionally generate the training toward alternative solutions that are genuinely distinct from the RN AdS geometry, yet reproduce the same boundary spectral data (Fig.~\ref{fig:Fixed_a}). Complementary results obtained with an interpretable machine-learning model, in place of the Neural ODE, are provided in Appendix~\ref{app2}.

Nonetheless, among the family of solutions consistent with $a=0$, generic training predominantly converges to the exact RN AdS geometry rather than to one of the alternative degenerate profiles. This suggests that the RN metric constitutes a smoother, more numerically favorable attractor in the Neural ODE optimization landscape.

This degeneracy can be understood from the structure of the low-energy fermionic excitations. Although the full bulk metrics differ from the AdS boundary to the horizon, their near-horizon geometries flow to the same $AdS_2 \times \mathbb{R}^2$ quantum critical region. Introducing the near-horizon radial coordinate $\zeta := 1 - z$ and expanding the metric ansatz \eqref{eq:nn_ansatz_At} as $\zeta \to 0$ brings Eq.~\eqref{eq:metric_ansatz} to the canonical near-horizon form:
\begin{equation}\label{eq:AdS2_metric}
    \mathrm{d} s^2 = -\frac{\zeta^2}{L_2^2} \, \mathrm{d} t^2 + \frac{L_2^2}{\zeta^2} \, \mathrm{d} \zeta^2 + L_x^2 \,\mathrm{d} \vec{x}_i^2 \,,
\end{equation}
with the effective $AdS_2$ radius $L_2$ and the $\mathbb{R}^2$ scale $L_x$ fixed by both the slope $a$ and the horizon values of the trained networks as
\begin{equation}\label{eq:AdS2_radius}
    L_2^2 = \frac{1}{3 + a + D_f(1;\theta_f)} \,, \quad L_x^2 = 1 + a + D_h(1;\theta_h) \,.
\end{equation}

For a massless probe fermion in the zero temperature limit, the low-energy scaling exponent $\nu_k$ and the IR conformal Green's function $\mathcal{G}_k(\omega)$ that govern the near-Fermi-surface behavior of $G_R(\omega,k)$ depend on the bulk metric \emph{only} through the two near-horizon combinations $L_2^2/L_x^2$ and $e_d^2$: Eq.~\eqref{eq:IR_exponent}~\cite{Faulkner:2010zz,Faulkner:2009wj}. Consequently, any two bulk profiles sharing the same AdS boundary asymptotics and the same near-horizon values of $L_2^2/L_x^2$ and $e_d^2$ are guaranteed to produce identical low-energy spectral functions near the Fermi surface, regardless of how they differ at intermediate bulk radius profile.

We verify this mechanism directly on our trained geometries: the solution of Fig.~\ref{fig:RN_trained_a} has $L_x^2/L_2^2=5.842$ and $1/e_d^2=11.75$, while the three degenerate solutions of Fig.~\ref{fig:Fixed_a} have $L_x^2/L_2^2=5.978$, $6.162$, $5.857$ and $1/e_d^2=12.14$, $11.79$, $12.32$, respectively --- all consistent, within numerical error, with the exact extremal RN AdS values $L_x^2/L_2^2=6$ and $1/e_d^2=12$. This confirms that it is the near-horizon data alone, and not the full radial profile, that controls the boundary spectral function near the Fermi surface: the boundary data fixes the local IR quantum critical universality class but leaves the UV-completed bulk spacetime that flows to it undetermined.

We emphasize an important caveat regarding this interpretation. The metric ansatz Eq.~\eqref{eq:nn_ansatz_At} is not required to satisfy the Einstein-Maxwell equations of motion for any specific bulk gravitational action; its functional form is fixed only by the AdS boundary conditions, extremal horizon regularity, and the fit to the boundary fermionic spectral data through the probe Dirac flow equation Eq.~\eqref{eq:dirac_flow_fh_explicit}. The isospectral degeneracy we observe is therefore a degeneracy of bulk metrics as sources of the probe-fermion response function, not a statement that these metrics are degenerate on-shell solutions of a fixed bulk equation of motion sourced by a definite gravitational action. Whether this degeneracy survives once the metric is required to self-consistently solve the Einstein-Maxwell equations with back-reaction from the fermion (or a bulk matter sector) is a distinct and interesting question that we leave for future work.

This isospectral non-identifiability is precisely the bulk-metric degeneracy expected on general holographic grounds at zero temperature, and it also constitutes a nontrivial success from a machine-learning standpoint: independent training runs, with no built-in knowledge of this expected universality, spontaneously converge onto distinct bulk profiles that nonetheless share the same near-horizon data --- indicating that the deep neural network isolates the physical IR fixed point rather than merely overfitting a single UV completion.

%%%%%%%%%%%%%%%%%%%%%%%%%%%
%
%%%%%%%%%%%%%%%%%%%%%%%%%%%
\section{Conclusions}\label{sec6}
In this work, we developed a physics-informed deep learning architecture based on Neural Ordinary Differential Equations (Neural ODEs) to tackle the holographic inverse problem for fermionic observables. By parameterizing unknown bulk fields with continuous deep neural networks and embedding the exact bulk Dirac equations into the backpropagation path, our framework can construct the emergent bulk spacetime and gauge potentials directly from boundary fermionic spectral functions.

Applying this framework to zero-temperature ($T=0$) extremal Reissner-Nordström AdS spacetime, we demonstrated successful reconstruction of the bulk geometry and gauge potential across three distinct quantum critical spectral regimes at fixed $U(1)$ probe fermion charge $q$: a non-Fermi liquid without quasiparticle excitations ($q=1$), a marginal Fermi liquid characteristic of strange-metal transport ($q=1.56$), and a sharp Fermi-liquid-like quasiparticle state ($q=2$). All three reconstructions achieved high fidelity, confirming that the framework is versatile and robust against quantitatively different boundary excitations. In Appendix~\ref{app1}, we further reconstruct a three-dimensional (BTZ) black hole background, showing that our deep learning framework is not restricted to a particular bulk dimensionality.

Furthermore, we expanded the scope of the inversion by simultaneously extracting the bulk geometry and a matter-sector parameter: promoting the probe Dirac charge $q$ to a trainable parameter alongside the network weights of bulk fields, using the marginal Fermi liquid data, the optimization converged to $q_{\mathrm{trained}}=1.5610$ (a $0.06\%$ relative deviation from the target $q=1.56$), with a final loss of $\mathcal{L}_{\mathrm{final}}=7.66\times10^{-6}$. This shows that the boundary fermionic spectral function encodes enough physical information to disentangle the background gravitational geometry from the parameters of the matter fields propagating on it.

We also examined the uniqueness of the holographic reconstruction by relaxing the near AdS boundary constraint of bulk spacetime, and found a geometrical degeneracy: bulk metric profiles that differ substantially throughout the radial bulk can reproduce the same boundary spectral data whenever they share the same near-horizon $AdS_2\times\mathbb{R}^2$ data. This is holographically expected, since for massless probe fermions the low-energy exponent $\nu_{k_F}$ and self-energy $\Sigma(\omega,k)$ depend only on the near-horizon geometry, not on the precise UV-complete bulk profile. From a machine-learning standpoint, the spontaneous emergence of such degenerate solutions across independent training runs is itself evidence that the network isolates this IR universality class, subject to the caveats on probe-limit validity and finite-grid resolution discussed in Sec.~\ref{sec5b}.

Altogether, our results establish a purely data-driven route from boundary fermionic spectra---comparable to angle-resolved photoemission spectroscopy (ARPES) experiments---to an emergent higher-dimensional holographic gravitational dual. A natural next step is to relax the zero-temperature assumption: at finite $T$ the Fermi surface is smeared out, so that the retarded Green's function (i.e., spectral functions) no longer develops a sharp pole at a well-defined momentum, and a recent study~\cite{Xiao:2026dxe} has already begun exploring neural-network reconstruction of charged AdS spacetimes directly from finite-temperature fermionic spectral functions. Extending our Neural ODE inverse framework to non-extremal black holes, and ultimately to real ARPES datasets from high-$T_c$ cuprates and heavy-fermion materials, is a promising direction for future work.

More broadly, our results speak to a central question facing the emerging program of data-driven holographic condensed matter theory: to what extent can a bulk gravitational dual be inferred directly from boundary data, without assuming a specific bulk action in advance? Physics-informed machine learning provides the technical means to pose this inverse problem in a controlled way, but our results show that any such program may also confront the intrinsic non-uniqueness of the reconstruction --- here traced to IR $AdS_2\times\mathbb{R}^2$ universality --- rather than treating a single fitted geometry as the unique gravitational dual. Identifying which features of an inferred bulk geometry are physically robust, and which are artifacts of this residual degeneracy, will be essential as these methods are applied to genuine experimental spectra.

%%%%%%%%%%%%%%%%%%%%%%%%%%%
%
%%%%%%%%%%%%%%%%%%%%%%%%%%%
\acknowledgments
We would like to thank {Yongjun Ahn, Johanna Erdmenger, Ki-Seok Kim, René Meyer, and Yunseok Seo} for valuable discussions and correspondence. 
HSJ was supported by an appointment to the JRG Program at the APCTP through the Science and Technology Promotion Fund and Lottery Fund of the Korean Government. HSJ was also supported by the Korean Local Governments -- Gyeongsangbuk-do Province and Pohang City.
This work was supported by the Basic Science Research Program through the National Research Foundation of Korea (NRF) funded by the Ministry of Science, ICT $\&$ Future Planning (NRF-2021R1A2C1006791) and the AI-based GIST Research Scientist Project grant funded by the GIST in 2025. This work was also supported by Creation of the Quantum Information Science R$\&$D Ecosystem (Grant No. 2022M3H3A106307411) through the National Research Foundation of Korea (NRF) funded by the Korean government (Ministry of Science and ICT).
The work of K.~H.~was supported in part by JSPS KAKENHI Grant No.~JP22H05111 and JP22H05115.
The work of DT was supported by RIKEN Special Postdoctoral Researchers Program.
All authors contributed equally to this paper and should be considered as co-first authors.

%%%%%%%%%%%%%%%%%%%%%%%%%%%
%    
%%%%%%%%%%%%%%%%%%%%%%%%%%%
\appendix

%%%%%%%%%%%%%%%%%%%%%%%%%%%
%    Added by Koji
%%%%%%%%%%%%%%%%%%%%%%%%%%%
\section{Interpretable machine learning of the bulk spacetime}\label{app2}
In this paper we have used the Neural ODE method to produce the bulk metric and the gauge field profiles. In general, once neural network methods are used, the interpretability of the solution profiles is lost, since the functional form of any neural network solution is made of multiply nested structure of activation functions, except for the case of Kolmogorov-Arnold networks \cite{liu2025kan}. This loss of interpretability is a major reason of failure of further reconstruction of the bulk {\it action} from the obtained reconstructed bulk profiles of the gravity and gauge fields. Thus here in this appendix we study a different method of machine learning with interpretable profiles.

To this end, the simplest method is to consider bulk profiles expanded in physically interpretable bases, 
\begin{equation}
    D_c(z) \equiv \sum_{n=0}^{K}\theta^{(c)}_n P_n(2z-1), \qquad c\in \{ f, h, A\}
    \label{eq:DLe}
\end{equation}
where $P_n$ is the Legendre function, $\theta^{(c)}_n$ is a real trainable coefficient, and $D_c(z)$ is the unknown function of \eqref{eq:nn_ansatz_At}. We put $a=0$ in this appendix. This Legendre expansion could be thought of physically as a higher derivative expansion of the bulk action, because the Legendre equation is a sort of mixture of the Taylor expansion and Fourier expansion thus its label $n$ can allow the physical interpretation of the radial momentum. $P_n(2z-1)$ is an eigenfunction of the operator $-(d/dz)[z(1-z)(d/dz)]$ with the eigenvalue $n(n+1)$, and it forms an orthonormal basis. The first $P_0$ is a constant, thus accommodates the RN solution
\begin{equation}
D^{\rm RN}_f = 3, \quad
D^{\rm RN}_h = 0,  \quad
D^{\rm RN}_A = 0.
\end{equation}
This Legendre expansion may help physical interpretation of the bulk profile when the machine learning is applied not just to reproduce the RN solution but with some actual material data. Together with some regularization term which suppresses large $n$ contributions, such as 
${\cal L}_{\rm Legendre} = n^2$, the obtained bulk profile incorporates consistency with the low energy expansion, meaning that the bulk action is a low energy effective gravity --- no need for strange higher derivative terms.

\begin{figure*}[t]
    \centering
    \includegraphics[width=0.7\textwidth]{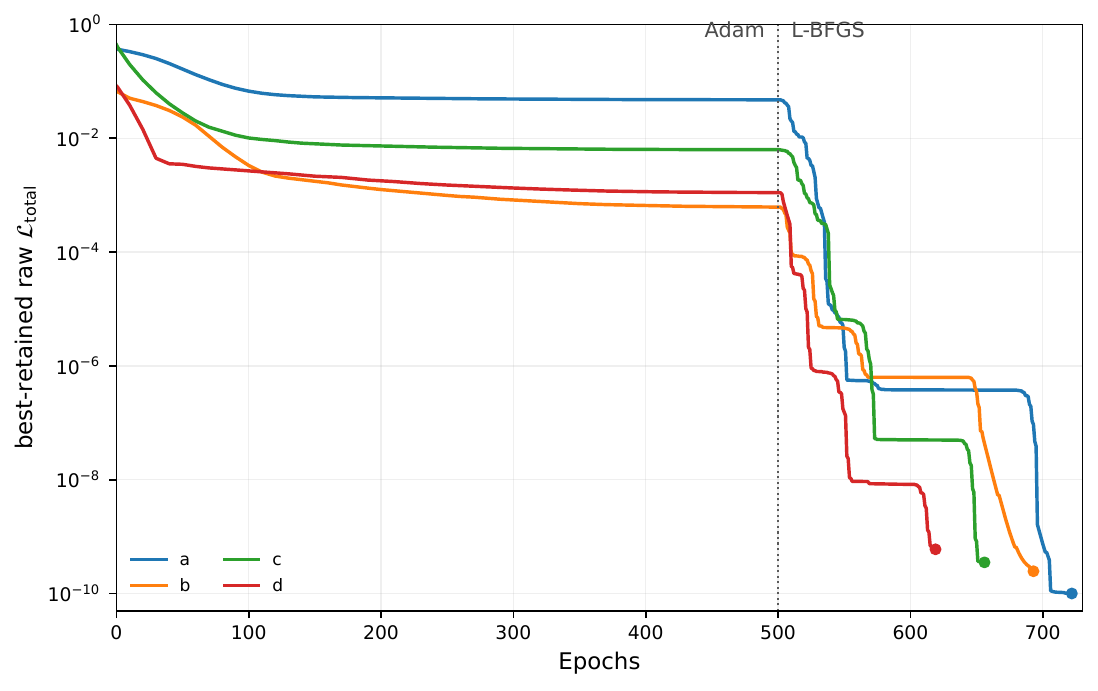}
\caption{The epoch evolution of the training for the $K=2$ case.}\label{fig:Legendre_los}
\end{figure*}

In this appendix, we just study the reproduction scheme of the RN solution, to look concretely at the degeneracy problem. (We do not use the loss ${\cal L}_{\rm Legendre}$ mentioned above.) Let us train the functions \eqref{eq:DLe} with the Data B: 
$q=1.56$, and the data points consist of the imaginary part of the Green's function evaluated at the direct product of $\omega\in \{-0.08,-0.04,0.04,0.08\}$ and $k\in\{0.65,0.85,1.05,1.25\}$.
So the data just consists of 16 points.
This number is important in judging whether we expect any degeneracy in the bulk profiles after the training, because for $K\leq 5$ the number of parameters $3K$ is smaller than the number of data points, while for $K\geq 6$ the number of parameters is larger. This means that for the training with $K\geq 6$ we generically expect that we will not reproduce the RN solution, resulting in degenerate bulk profiles.\footnote{The degeneracy issue is closely related to the linearity of the ODE of concern. Note that the equation \eqref{eq:dirac_flow_fh_explicit} which we integrate in the bulk looks nonlinear (in $\xi$) but it is in fact linear in fields ($y_\pm$ and $z_\pm$). This means that the integrated results of the equation are degenerate and depend only on the path-ordered exponential of the matrix kernel defined in \eqref{eq:y_equation}. What makes the integration nonlinear is $\omega$ and $k$ which appear as a multiplication in \eqref{eq:y_equation}, thus intrinsically the number of data points in the current setup is the one which determines the degrees of freedom of the issue.
} However, as we have discussed in the main text, we have actually encountered a degeneracy due to the fact that the data only depends on the bulk metric profile of the IR end. This means that even in this Legendre interpretable machine learning with $0<K\leq 5$, we expect degenerate bulk solutions.

The loss function is arranged as
\begin{align}
 {\cal L}_{\mathrm{total}}
 &\equiv {\cal L}_{\mathrm{data}}+10^6\mathcal L_{\mathrm{dom}},
 \\
 \mathcal L_{\rm data}
 &\equiv \frac{N^{-1}\sum_{i=1}^{N}
 [{\cal A}_{\boldsymbol\theta}(\omega_i,k_i)-{\cal A}_i^{\mathrm{data}}]^2}
 {\max\{16^{-1}\sum_{i=1}^{16}({\cal A}_i^{\mathrm{data}})^2,10^{-4}\}},
 \nonumber\\
 \mathcal L_{\mathrm{dom}}
 &\equiv \frac1{257}\sum_{j=1}^{257}
 \left[\operatorname{ReLU}(5\times10^{-3}-(1+2z_j+z_j^2D_f(z_j)))^2
 \right.\nonumber \\
 & \qquad \left.+\operatorname{ReLU}(5\times10^{-3}-h(z_j))^2\right].
\end{align}
The first term is for the data fitting, while the second term is to make sure that the functions $f(z)$ and $h(z)$ are positive, and we picked up evaluation points $z_j\equiv (j-1)/256$ (with $j=1,2,\cdots, 257$).

The training history for the various initial conditions for the parameters $\theta$ in the case $K=2$ is shown in Fig.~\ref{fig:Legendre_los}. All trainings attained
lowered loss values of ${\cal O}(10^{-8})$. 

\begin{figure*}[t]
    \centering
    \includegraphics[width=1\textwidth]{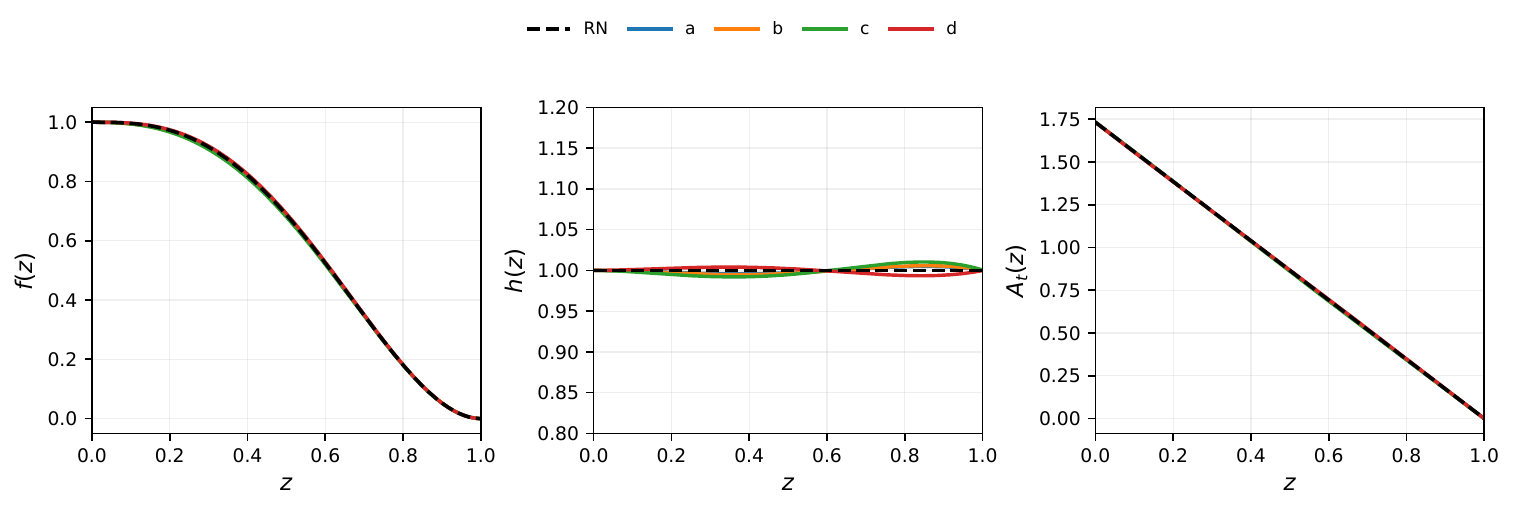}
    \includegraphics[width=1\textwidth]{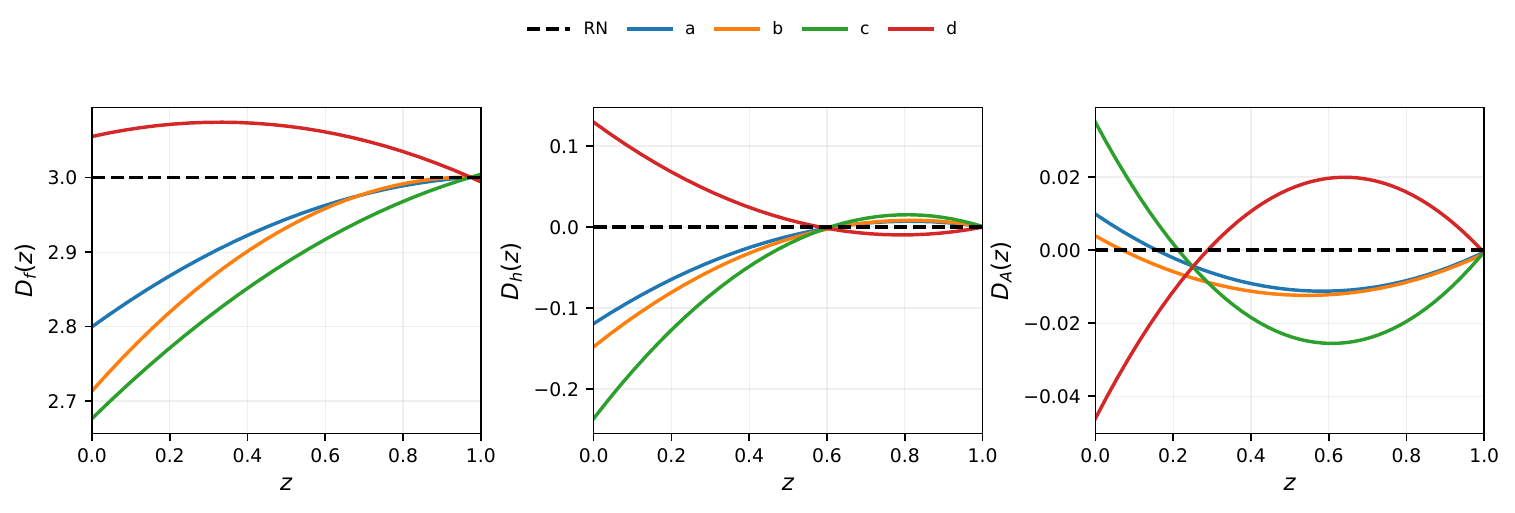}
\caption{The obtained bulk profiles (top) and the functions $D_c(z)$ (bottom) for the case $K=2$, with various initial values of $\theta$ for the training. Although the trained bulk profiles (top) look the same as the RN solution, they are indeed different, as obviously shown in the bottom figures shown in $D_c(z)$ variables.}\label{fig:Legendre}
\end{figure*}

The trained bulk profiles are shown in Fig.~\ref{fig:Legendre}. It provides bulk configurations looking quite different from the RN solution, keeping the Green function data intact. This means the degeneracy, which has been expected from the observation that the data actually looks at the bulk profile only at the IR end (which is the extremal black hole horizon).

In fact, as seen in the bottom right panel of Fig.~\ref{fig:Legendre}, all the trained profiles go to $D_f=3$ at the horizon $z=1$. This is consistent with the fact that the data is solely made out of the IR AdS${}_2$ geometry with \eqref{eq:AdS2_radius}.

This same degeneracy is also found, albeit less frequently, in the Neural ODE framework of Sec.~\ref{sec5b}: standard training at $a=0$ reliably converges to the exact RN AdS solution (Sec.~\ref{sec5}), but atypical initial conditions occasionally drive the optimization to alternative solutions. Three such solutions, shown in Fig.~\ref{fig:Fixed_a}, differ substantially both from the exact RN AdS geometry and from one another, yet reproduce the same boundary spectral data, with final losses of $\mathcal{L}_{\rm data}=1.65\times10^{-5}$, $5.92\times10^{-4}$, and $5.05\times10^{-4}$, respectively.

In actual situations with experimental datasets for the fermion Green's functions, it is better to use large $K$ Legendre profiles but with the loss function ${\cal L}_{\rm Legendre} = n^2$ to suppress the large radial momenta in the bulk. This will naturally accommodate both the expressibility of the profile functions and the resolution of the degeneracy to obtain a physical bulk configuration which allows low energy effective action in the bulk. The goal of the reconstruction program is not just reproducing the bulk metric profiles but goes to pin down the bulk action itself, which should be derivative expanded to make sure the low energy effective description of the holographic gravitational picture.

We end with some comments on the interpretable neural networks which have been developed in \cite{Hashimoto:2018ftp} and subsequent papers. There the bulk is directly interpreted as a neural network, and the neural network weights are the bulk metrics. This method actually allows the direct interpretation of the neural network weights, but it suffers from the same degeneracy problem: when the number of data points is small, the bulk metric has a lot of degeneracy. In particular, to make the bulk propagation numerically trusted, one needs to introduce a lot of bulk radial points on each of which the metric function provides the weights and thus the number of network parameters gets huge, resulting in the degeneracy. For example in \cite{hashimoto2025machine} the Runge-Kutta layers were introduced to make the bulk integration stable, with a large number of layers. This method works only when the number of data points is comparably large. In realistic cases with experimental data at nonzero temperature, we are short of the data points, meaning that the direct interpretable neural network such as the one in \cite{Hashimoto:2018ftp} may not be suitable to use.

%%%%%%%%%%%%%%%%%%%%%%%%%%%
%    
%%%%%%%%%%%%%%%%%%%%%%%%%%%
\section{Emergent spacetime from analytic fermionic spectral functions}\label{app1}
As another benchmark of the bulk spacetime reconstruction from boundary fermionic spectral function, we consider three-dimensional black hole geometry, especially Bañados-Teitelboim-Zanelli (BTZ) black holes~\cite{Banados:1992wn,Banados:1992gq}. Since the retarded Green's function for fermions can be computed analytically in a BTZ black hole background~\cite{Iqbal:2009fd}, we utilize this analytic fermionic spectral function as input data for bulk reconstruction within our Neural ODE framework. This serves as a complementary approach to our numerical analysis presented in the main text.

\subsection{Analytic fermionic Green's functions}\label{appsec1}
The BTZ metric is given by
\begin{equation}\label{eq:btz_metric_r}
\dd s^2 = -\left(r^2-1\right)\dd t^2 +\frac{\dd r^2}{r^2-1} +r^2 \dd x^2 \,,
\end{equation}
where the horizon is at $r=1$, and Hawking temperature reads $T=1/(2\pi)$. Following~\cite{Iqbal:2009fd}, it is convenient to introduce
\begin{equation}\label{eq:btz_rho_coordinates}
r = \cosh\rho \,,
\end{equation}
for which the metric \eqref{eq:btz_metric_r} becomes
\begin{equation}\label{eq:btz_metric_rho}
\dd s^2= -\sinh^2\rho\, \dd t^2 +\cosh^2\rho\, \dd x^2+\dd \rho^2 \,.
\end{equation}
In this frame, the Fourier decomposition takes the form
\begin{gather}
    \Psi
    =
    e^{-i\omega t+ikx}\psi(\rho).
    \label{eq:btz_momenta}
\end{gather}

Starting from the Dirac equation discussed in Sec.~\ref{sec2}, its radial part takes the form
\begin{equation}\label{eq:btz_dirac_rho}
\begin{aligned}
\bigg[
\Gamma^{\underline{\rho}}
\bigg(\partial_\rho
&+\frac{1}{2}
\left(\frac{\cosh\rho}{\sinh\rho}+\frac{\sinh\rho}{\cosh\rho}\right)\bigg) \\
&+ i\left(\frac{\Gamma^{\underline{x}}k}{\cosh\rho}-\frac{\Gamma^{\underline{t}}\omega}{\sinh\rho}\right)-m\bigg]\psi=0 \,,
\end{aligned}
\end{equation}
where
\begin{equation}\label{eq:btz_gamma_matrices}
\Gamma^{\underline{\rho}}=\sigma_3 \,,
\quad
\Gamma^{\underline{t}}=i\sigma_2 \,,
\quad
\Gamma^{\underline{x}}=\sigma_1 \,,
\quad
\psi^{\mathrm T}=(\psi_+,\psi_-) \,.
\end{equation}
To reduce Eq.~\eqref{eq:btz_dirac_rho} to a
hypergeometric system, we define
\begin{equation}\label{eq:btz_chi_definition}
\psi_\pm = \sqrt{\frac{\cosh\rho\pm\sinh\rho}{\cosh\rho\,\sinh\rho}}\left(\chi_1\pm\chi_2\right)\,, \quad
u=\tanh^2\rho \,.
\end{equation}
The compact radial coordinate $u$ places the horizon at
$u=0$ and the AdS boundary at $u=1$. The two first-order equations then become
\begin{align}\label{eq:btz_chi1_exact}
\begin{split}
2(1-u)\sqrt{u}\,\,\partial_u\chi_1 & - i\left(\frac{\omega}{\sqrt{u}}+k\sqrt{u}\right)\chi_1 \\
&=\left[m-\frac{1}{2}+i(\omega + k)\right]\chi_2 \,,\\
2(1-u)\sqrt{u}\,\,\partial_u\chi_2&+i\left(\frac{\omega}{\sqrt{u}} + k\sqrt{u}\right)\chi_2 \\
&=\left[m-\frac{1}{2}-i(\omega + k)\right]\chi_1 \,.
\end{split}
\end{align}
\begin{figure*}[t]
\centering
\includegraphics[width=0.35\textwidth]{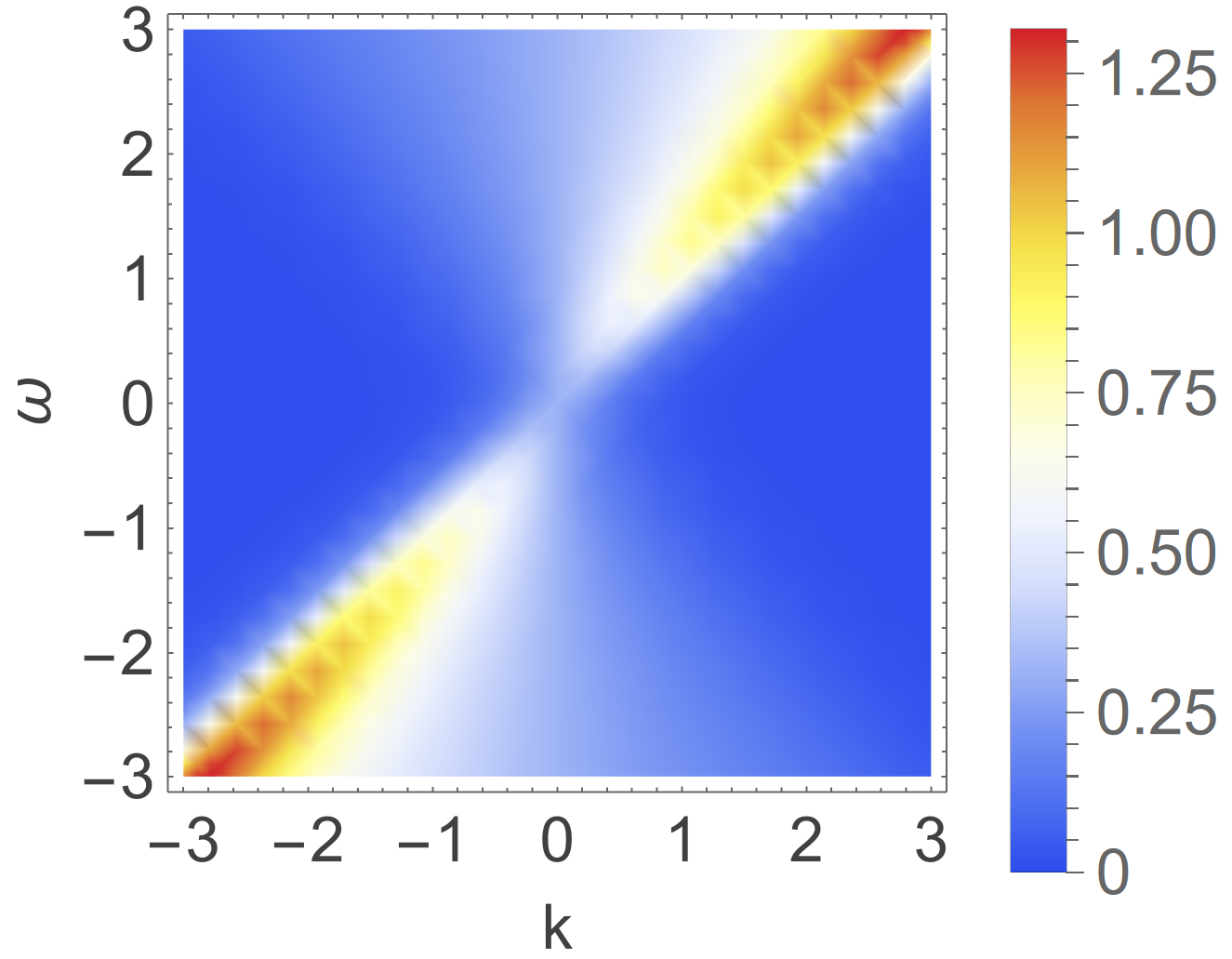}
\qquad
\includegraphics[width=0.5\textwidth]{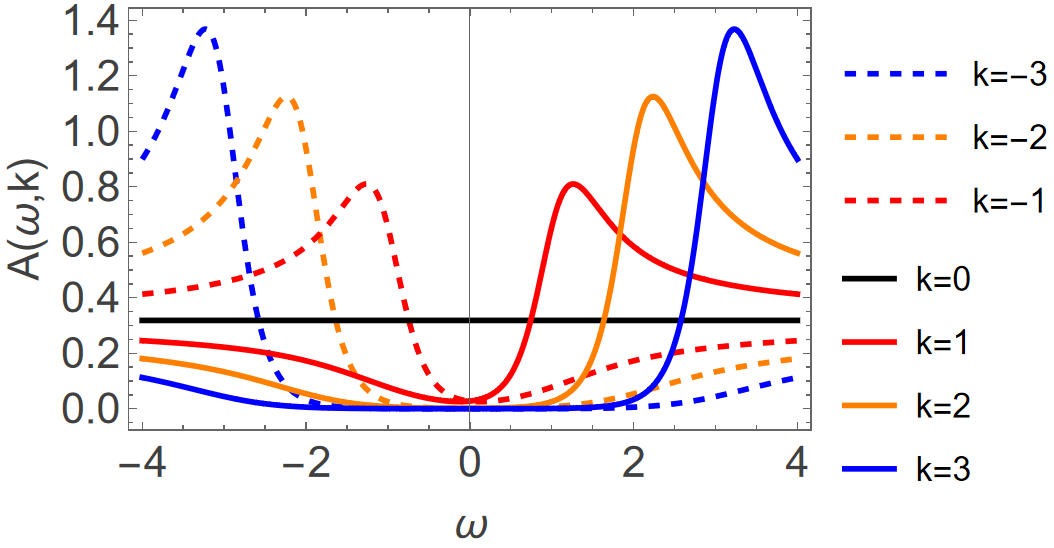}
\caption{The analytic fermionic spectral function of the BTZ black hole for $m=0$. The left panel displays the density plot in the $(\omega,k)$ plane. The right panel shows $\omega$-dependent slices at fixed momenta.}\label{fig:btz_spectral}
\end{figure*}
These Dirac equations can be solved analytically with the hypergeometric function ${}_2F_1(\mathfrak{a},\mathfrak{b};\mathfrak{c};u)$ for the infalling solution as follows:
\begin{align}\label{eq:btz_chi2_solution}
\begin{split}
\chi_1(u) &= \frac{\mathfrak{a}-\mathfrak{c}}{\mathfrak{c}} u^{\alpha+\frac{1}{2}}(1-u)^\beta{}_2F_1(\mathfrak{a},\mathfrak{b}+1;\mathfrak{c}+1;u) \,, \\
\chi_2(u) &= u^\alpha(1-u)^\beta{}_2F_1(\mathfrak{a},\mathfrak{b};\mathfrak{c};u) \,,
\end{split}
\end{align}
where
\begin{equation}\label{eq:btz_alpha_beta}
\begin{gathered}
\alpha=-\frac{i\omega}{2} \,, \qquad \beta=-\frac{1}{4}+\frac{m}{2} \,, \\
\mathfrak{a} = \frac{1}{2}\left(m+\frac{1}{2}\right) -\frac{i}{2}(\omega - k) \,, \\
\mathfrak{b} = \frac{1}{2}\left(m-\frac{1}{2}\right) -\frac{i}{2}(\omega + k) \,, \qquad
\mathfrak{c} = \frac{1}{2}-i\omega \,.
\end{gathered}
\end{equation}
Plugging \eqref{eq:btz_chi2_solution} into \eqref{eq:btz_chi_definition}, one can find the near AdS boundary behaviors as
\begin{align}\label{eq:btz_psi_plus_boundary}
\begin{split}
\psi_+ & \approx A(1-u)^{\frac{1}{2}-\frac{m}{2}} + B(1-u)^{1+\frac{m}{2}} \,, \\
\psi_- & \approx C(1-u)^{1-\frac{m}{2}} + D(1-u)^{\frac{1}{2}+\frac{m}{2}} \,.
\end{split}
\end{align}
For $m\geq 0$, $A$ is the source and $D$ is the response in the standard quantization.  With the overall sign chosen consistently with the positive spectral-density convention in Eq.~\eqref{eq:spectral_function}, the retarded Green's function is
\begin{align}\label{eq:btz_analytic_green}
\begin{split}
G_R(\omega,k) &= - i\frac{D}{A} = i \frac{\Gamma\left(\frac{1}{2}-m\right)} {\Gamma\left(\frac{1}{2}+m\right)} \mathcal{F}_L(\omega,k) \mathcal{F}_R(\omega,k) \,,
\end{split}
\end{align}
where
\begin{align}\label{eq:btz_left_factor}
\begin{split}
\mathcal{F}_L(\omega,k) &= \frac{\Gamma\left(\frac{1}{4} + \frac{m}{2} - i(\omega-k) \right)}{\Gamma\left(\frac{3}{4} - \frac{m}{2} - i(\omega-k)\right)} \,,\\
\mathcal{F}_R(\omega,k)&=\frac{\Gamma\left(\frac{3}{4} + \frac{m}{2} - i(\omega+k)\right)}{\Gamma\left(\frac{1}{4} - \frac{m}{2} - i(\omega+k)\right)} \,.
\end{split}
\end{align}
We display the fermionic spectral function \eqref{eq:btz_analytic_green} for massless fermions in Fig. \ref{fig:btz_spectral}.

\subsection{Neural ODE reconstruction}\label{appsec2}
Next, using the Neural ODE framework, we reconstruct the BTZ geometry from the analytic spectral function given in Fig.~\ref{fig:btz_spectral}. To align with the inverse-problem framework used in the main text, we rewrite the Dirac equation in terms of the new compact coordinate $u=1 - z^2$, where the metric ansatz~\eqref{eq:metric_ansatz} becomes
\begin{align}\label{eq:btz_metric_u_general}
\dd s^2 &= -\frac{f(u)}{1-u}\,\dd t^2 + \frac{\dd u^2}{4f(u)(1-u)^2} +\frac{h(u)}{1-u}\,\dd x^2 \,.
\end{align}
Here, the horizon and AdS boundary are located at $u=0$ and
$u=1$, respectively. The black hole horizon and asymptotic AdS
conditions are
\begin{equation}\label{eq:btz_metric_conditions}
f(0)=0 \,, \qquad f(1)=1 \,, \qquad h(1)=1 \,.
\end{equation}
In this new coordinate system, the exact BTZ solution \eqref{eq:btz_metric_r} becomes 
\begin{equation}\label{eq:btz_exact_profiles}
f(u)=u \,, \qquad h(u)=1 \,.
\end{equation}

Substituting Eq.~\eqref{eq:btz_metric_u_general} into the
Dirac equation and using the $\chi_{1,2}$ basis defined in
Eq.~\eqref{eq:btz_chi_definition}, one can find
\begin{align}\label{eq:btz_general_chi1}
\begin{split}
4(1-u)\chi_1'(u) &+ \left[ \mathcal{P}(u)-i\mathcal{Q}(u) \right]\chi_1(u) \\
&\qquad  = \left[\mathcal{R}(u)+i\mathcal{S}(u)\right]\chi_2(u) \,, \\
4(1-u)\chi_2'(u) &+ \left[\mathcal{P}(u)+i\mathcal{Q}(u)\right]\chi_2(u)\\
&\qquad  = \left[ \mathcal{R}(u)-i\mathcal{S}(u) \right]\chi_1(u) \,,
\end{split}
\end{align}
where 
\begin{align}\label{eq:btz_P}
\begin{split}
\mathcal{P}(u) &=(1-u)\left[-\frac{1}{u}+\frac{f'(u)}{f(u)}+\frac{h'(u)}{h(u)} \right] \,,\\
\mathcal{Q}(u)&=2\left[\frac{\omega}{f(u)}+\frac{k\sqrt{u}}{\sqrt{f(u)}\sqrt{h(u)}}\right] \,, \\
\mathcal{R}(u)&=-\frac{1}{\sqrt{u}}+\frac{2m}{\sqrt{f(u)}} \,,\\
\mathcal{S}(u)&=2\left[\frac{\omega\sqrt{u}}{f(u)}+\frac{k}{\sqrt{f(u)}\sqrt{h(u)}} \right]\,.
\end{split}
\end{align}
\begin{figure*}[t]
    \centering
    \includegraphics[width=0.35\textwidth]{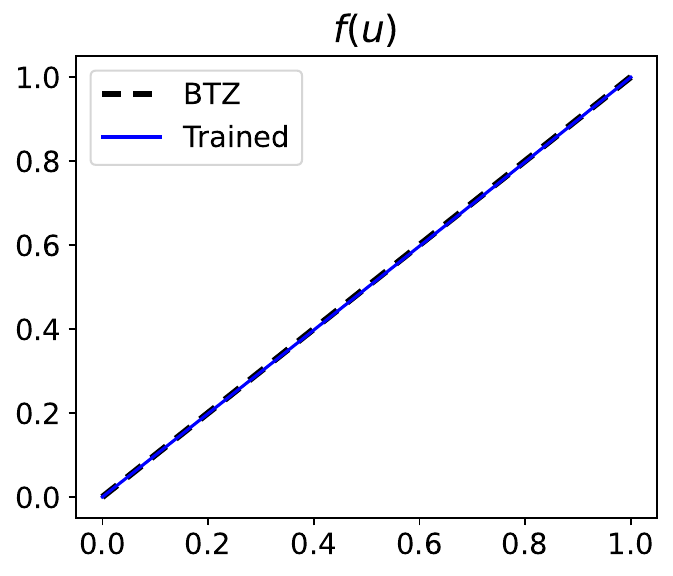}
\qquad
    \includegraphics[width=0.35\textwidth]{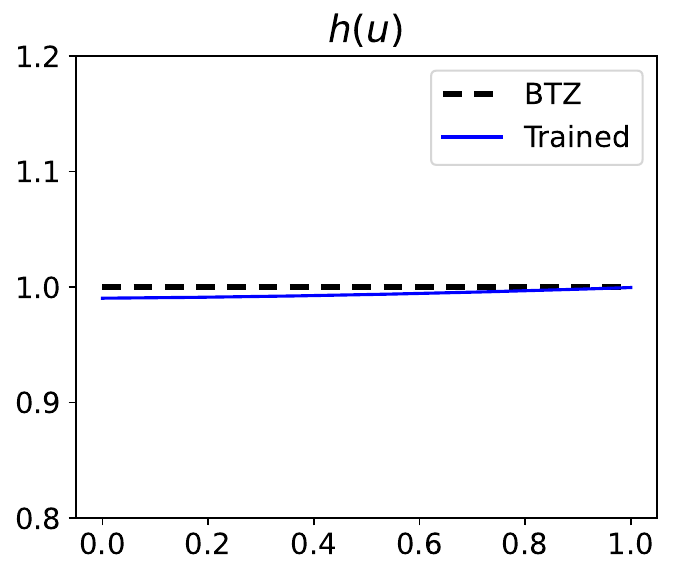}
\caption{Neural ODE reconstruction of the BTZ black hole spacetime from the massless fermionic spectral data (Fig. \ref{fig:btz_spectral}). The trained metric functions are in good agreement with the reference functions of the exact BTZ background \eqref{eq:btz_exact_profiles}.}\label{fig:btz_reconstruction}
\end{figure*}
The infalling behavior at the horizon and the leading falloff near the AdS boundary are made explicit via
\begin{align}\label{eq:btz_X1_definition}
\begin{split}
\chi_1(u) &= u^{-\frac{i\omega}{2f'(0)}} \sqrt{u}\, (1-u)^{-\frac{1}{4}+\frac{m}{2}} X_1(u) \,,\\
\chi_2(u)&=u^{-\frac{i\omega}{2f'(0)}}(1-u)^{-\frac{1}{4}+\frac{m}{2}} X_2(u) \,.
\end{split}
\end{align}
The functions $X_1$ and $X_2$ are regular at the horizon.
Near the AdS boundary $u=1$, they are expanded as
\begin{align}\label{eq:btz_X1_boundary}
\begin{split}
X_1(u) &\sim \sum_{n=0}a_n(1-u)^n + (1-u)^{\frac{1}{2}-m} \sum_{n=0}b_n(1-u)^n \,, \\
X_2(u) &\sim \sum_{n=0}c_n(1-u)^n + (1-u)^{\frac{1}{2}-m} \sum_{n=0}d_n(1-u)^n \,.
\end{split}
\end{align}
For each pair $(\omega_i,k_i)$, we solve the coupled equations \eqref{eq:btz_general_chi1} for $X_1$ and $X_2$ from the near-horizon region to the near-boundary region. The numerical solutions are then fitted to Eqs.~\eqref{eq:btz_X1_boundary} to determine $a_0,b_0,c_0$, and $d_0$. The Green's function can be expressed in terms of these coefficients as
\begin{equation}
G_R=
\begin{cases}
\displaystyle
-2i\, \frac{b_0+d_0}{a_0-c_0}\,, & m<0\,, \\[8pt]
\displaystyle
-\frac{i}{2}\,
\frac{a_0-c_0}{b_0+d_0} \,, & m\geq 0 \,.
\end{cases}\label{eq:btz_fitted_green}
\end{equation}

We use the same neural-network
architecture, ODE solver, Adam--L-BFGS optimization
procedure as in
Sec.~\ref{sec4}. The two changes specific to the
BTZ reconstruction are the hard constraints
\begin{equation}\label{eq:btz_nn_ansatz}
f_\theta(u)= u \, D_f(u;\theta_f) \,,\quad
h_\theta(u)= D_h(u;\theta_h) \,,
\end{equation}
and the inclusion of the AdS boundary conditions \eqref{eq:btz_metric_conditions} in the
loss function for computational convenience:
\begin{equation}\label{eq:btz_condition_loss}
\mathcal{L}_{\mathrm{cond}} = \left[f_\theta(1)-1\right]^2 + \left[ h_\theta(1)-1\right]^2 \,.
\end{equation}
The total loss is therefore
\begin{equation}\label{eq:btz_total_loss}
\mathcal{L} = \mathcal{L}_{\mathrm{data}} + \mathcal{L}_{\mathrm{cond}} \,,
\end{equation}
where $\mathcal{L}_{\mathrm{data}}$ has the same form as
Eq.~\eqref{eq:spectral_loss}. 

The analytic massless fermionic spectral function shown in Fig.~\ref{fig:btz_spectral} is sampled on an approximately regular grid, following the same digitization procedure used for the spectral data in the main text. The reconstruction displayed below uses the window $0<\omega<1$ and $0<k<1$.

As shown in Fig.~\ref{fig:btz_reconstruction}, the trained functions are consistent with the analytic BTZ spacetime \eqref{eq:btz_exact_profiles}. The optimization achieves a final loss of $8.03 \times 10^{-4}$. The successful recovery of both the blackening function $f(u)$ and the spatial metric component $h(u)$ from the boundary fermionic spectrum provides an additional benchmark for the black hole spacetime reconstruction from the boundary fermionic data.

%%%%%%%%%%%%%%%%%%%%%%%%%%%
%    
%%%%%%%%%%%%%%%%%%%%%%%%%%%
\providecommand{\href}[2]{#2}\begingroup\raggedright\endgroup

\end{document}